\documentclass[]{aa}
\makeatletter
\renewcommand*\aa@pageof{, page \thepage{} of \pageref*{LastPage}}
\makeatother
\newcommand{\cha}{\textit{Chandra\/}}

\def\flu{{erg\,s$^{-1}$\,cm$^{-2}$}}

\def\leg{{\em COSMOS-Legacy}\/}

\def\aap{A\&A}

\def\apjl{ApJL}
\def\apjs{ApJS}

\usepackage{natbib}
\usepackage{float}
\usepackage{color}
\usepackage{graphicx}
\usepackage{textcomp,gensymb}
\usepackage{array}
\usepackage{mathtools}
\usepackage{multirow,makecell}
\usepackage{enumitem}
\usepackage{longtable}
\usepackage{hyperref}
\begin{document}

\title{Redshift identification of X-ray selected active galactic nuclei in the J1030 field: searching for large-scale structures and high-redshift sources
}
\titlerunning{Spec- and photo-z of X-ray AGN in the J1030 field}

\author{Stefano~Marchesi\inst{1,2} \and Marco~Mignoli\inst{1} \and Roberto~Gilli\inst{1} \and  Alessandro~Peca\inst{3} \and Micol~Bolzonella\inst{1} \and Riccardo~Nanni\inst{4} \and Marianna~Annunziatella\inst{5} Barbara~Balmaverde\inst{6} \and Marcella~Brusa\inst{7,1} \and Francesco~Calura\inst{1} \and Letizia~P.~Cassarà\inst{8} \and Marco~Chiaberge\inst{9,10} Andrea~Comastri\inst{1} \and Felice~Cusano\inst{1} \and Quirino~D'Amato\inst{11,7} \and Kazushi~Iwasawa\inst{12,13} \and Giorgio~Lanzuisi\inst{1} \and Danilo~Marchesini\inst{14,15} \and Takahiro~Morishita\inst{16} \and Isabella~Prandoni\inst{11} \and Andrea~Rossi\inst{1} \and Paolo~Tozzi\inst{17} \and  Cristian~Vignali\inst{7,1} \and Fabio~Vito\inst{18} \and Giovanni~Zamorani\inst{1} \and Colin~Norman\inst{16}
}

\institute{INAF - Osservatorio di Astrofisica e Scienza dello Spazio di Bologna, Via Piero Gobetti, 93/3, 40129, Bologna, Italy 
\and Department of Physics and Astronomy, Clemson University, Kinard Lab of Physics, Clemson, SC 29634, USA 
\and Department of Physics, University of Miami, Coral Gables, FL 33124, USA 
\and Department of Physics, University of California, Santa Barbara, CA 93106-9530, USA 
\and Centro de Astrobiologia (CSIC-INTA), Ctra de Torrej\'{o}n a Ajalvir, km 4, E-28850 Torrej\'{o}n de Ardoz, Madrid, Spain 
\and INAF – Osservatorio Astrofisico di Torino, Via Osservatorio 20, 10025 Pino Torinese, Italy
\and Dipartimento di Fisica e Astronomia, Università degli Studi di Bologna, via Gobetti 93/2, 40129 Bologna, Italy 
\and INAF -- Istituto di Astrofisica Spaziale e Fisica Cosmica, Via Alfonso Corti 12, 20133, Milan, Italy
\and AURA for the European Space Agency (ESA), ESA Office, Space Telescope Science Institute, 3700 San Martin Drive, Baltimore, MD 21218, USA
\and  Johns Hopkins University - Center for Astrophysical Sciences, 3400 N. Charles Street, Baltimore, MD 21218, USA
\and INAF -- Istituto di Radioastronomia, via P. Gobetti 101, 40129 Bologna, Italy 
\and Institut de Ci\`encies del Cosmos (ICCUB), Universitat de Barcelona (IEEC-UB), Mart\'i i Franqu\`es, 1, 08028 Barcelona, Spain
\and ICREA, Pg. Lu\'is Companys 23, 08010 Barcelona, Spain 
\and Department of Physics and Astronomy, Tufts University, Medford, MA 02155, USA 
\and International Research Fellow of Japan Society for the Promotion of Science (Invitational Fellowships for Research in Japan — Short-term) 
\and Space Telescope Science Institute, 3700 San Martin Drive, Baltimore, MD 21218, USA 
\and INAF -- Osservatorio Astrofisico di Arcetri, Largo E. Fermi, I-50125 Firenze, Italy
\and Scuola Normale Superiore, Piazza dei Cavalieri 9, Pisa, Italy}

\abstract{We publicly release the spectroscopic and photometric redshift catalog of the sources detected with \cha\ in the field of the $z$=6.3 quasar SDSS J1030+0525. This is currently the fifth deepest extragalactic X-ray field, and reaches a 0.5--2\,keV flux limit $f_{\rm 0.5-2}$=6$\times$10$^{-17}$\,\flu.
By using two independent methods, we measure a photometric redshift for 243 objects, while 123 (51\,\%) sources also have a spectroscopic redshift, 110 of which coming from an INAF-Large Binocular Telescope (LBT) Strategic Program. We use the spectroscopic redshifts to determine the quality of the photometric ones, and find it in agreement with that of other X-ray surveys which used a similar number of photometric data-points. In particular, we measure a sample normalized median absolute deviation $\sigma_{\rm NMAD}$=1.48$\times$median(||$z_{\rm phot}$-$z_{\rm spec}$||/(1+$z_{\rm spec}$))=0.065.

We use these new spectroscopic and photometric redshifts to study the properties of the \cha\ J1030 field. We observe several peaks in our spectroscopic redshift distribution between $z$=0.15 and $z$=1.5, and find that the sources in each peak are often distributed across the whole \cha\ field of view. This confirms that X-ray selected AGN can efficiently  track large-scale structures over physical scales of several Mpc. Finally, we computed the \cha\ J1030 {\it z}$\,>\,$3 number counts: while the spectroscopic completeness at high-redshift of our sample is limited, our results point towards a potential source excess at $z\geq$4, which we plan to either confirm or reject in the near future with dedicated spectroscopic campaigns.
}


\maketitle

\section{Introduction}

It is now well established that the star formation (SF) in galaxies and black hole accretion in an ``active galactic nucleus'' (AGN) phase peak at redshifts $z\sim$2--3 \citep[e.g.,][]{madau14,ueda14,ananna19}, hinting at a co-evolutionary supermassive black hole (SMBH)-host growth scenario. However, the origin and physical causes of this scenario are still debated \citep[e.g.,][]{alexander12,fiore17,izumi19}

To fully understand this co-evolutionary process, one needs large samples of accreting supermassive black holes, with excellent multiwavelength coverage. In particular, X-ray surveys are among the most efficient methods to track the mass growth of SMBHs in the AGN phase over a wide range of redshifts and accretion efficiencies. In fact, X-rays are significantly less contaminated by non-AGN emission processes than optical and infrared surveys, and even in the deepest X-ray surveys the non-AGN population is subdominant, if not fully negligible \citep[e.g.,][]{donley08,donley12,lehmer12,marchesi16a,luo17}. 
This highlights the fact that X-rays are one of the best ways to detect intrinsically faint AGN, whose optical emission is dominated by non-nuclear processes and whose SED can therefore be fitted with a host galaxy template, with no AGN contribution \citep[see, e.g.,][]{marchesi16a}.
Even more importantly, X-ray surveys can detect a significant population of obscured and even Compton thick (i.e., with column density N$_{\rm H}\geq$10$^{24}$\,cm$^{-2}$) AGNs, up to redshift $z\sim$2--3 \citep[e.g.,][]{comastri11,georgantopoulos13,lanzuisi15,lanzuisi18} that are missed by optical surveys, where the SED is once again dominated by non-AGN related processes. 
Notably, obscured AGNs are expected to dominate the census of SMBHs at Cosmic Dawn, where the vast majority of SMBHs were highly efficiently accreting in a dense environment \citep[e.g.,][]{lanzuisi18,vito18,matsuoka19}.
Indeed, deep X-ray surveys have shown that the incidence of obscuration in AGN increases towards early cosmic times, likely because of the dense ISM measured in their hosts \citep{lanzuisi18,vito18,circosta19,damato20}. For example, the fraction of obscured QSOs with $L_{bol} \gtrsim 10^{46}$\,erg\,s$^{-1}$ rises from 10-20\% in the local Universe to 80-90\% at $z\sim$4 \citep{vito18}. Simulations support this increase, and further predict that the obscured QSO fraction may exceed 99\% at $z\gtrsim$6 \citep{ni20}. 

Among the different X-ray facilities currently available, the \cha\ X-ray telescope, with its excellent sub-arcsecond resolution, is the ideal instrument for deep surveys aimed at characterizing the intrinsically faint and/or heavily obscured AGN population, as shown in several works in the past ten years. Examples of deep, pencil-beam (i.e., covering an area $<$0.5\,deg$^2$) surveys are the 2\,Ms \cha\ Deep Field-North \citep[CDF-N][]{xue16}, the 7\,Ms \cha\ Deep Field-South \citep[CDF-S][]{luo17}, AEGIS-XD \citep[][]{nandra15}, the \cha\ UKIDSS Ultra Deep Survey \citep[X-UDS][]{kocevski18} and the SSA22 Survey \citep[][]{lehmer09}. On a larger area, comparable depths have been achieved by the \cha\ COSMOS-Legacy survey \citep[][]{civano16,marchesi16a}.

To fully exploit their efficiency in identifying a population of sources that would be missed by other facilities, however, X-ray surveys need to be complemented by extensive multi-wavelength information. In particular, the redshift completeness of the X-ray samples should ideally be as close as possible to 100\,\%. To do so, optically bright sources are targeted with spectroscopic campaigns \citep[see, e.g.,][for the CANDELS/COSMOS field]{lilly07,lilly09,kriek15}. For the remaining part of the sample, which is usually consists of those sources that are too faint to obtain a reliable spectroscopic redshift, photometric redshifts are computed, taking advantage of extensive imaging campaigns in the optical and near-infrared \citep[see, e.g.,][for the \cha\ Deep Field South; \citealt{ilbert09,salvato11,laigle16,marchesi16a} for the COSMOS field; \citealt{ananna17} for the Stripe 82X field]{hsu14}.

Recently, \cha\ targeted with a deep, $\sim$0.5\,Ms observation a 335\,arcmin$^2$ region centered on the $z$=6.31 quasar SDSS J1030+0525.  SDSS J1030+0525 is one of the first quasars detected at {\it z}$\,>\,$6 by the Sloan Digital Sky Survey \citep[SDSS][]{fan01}. This field is known to be highly biased, since it hosts both a galaxy overdensity at $z$=6.3, around the SDSS J1030+0525 \citep{morselli14,balmaverde17,mignoli20}, and another overdensity at $z$=1.7, around a Fanaroff-Riley type II (FRII) radio galaxy \citep{nanni18,gilli19,damato20}. The flux limit achieved by the \cha\ J1030 survey \citep{nanni20} makes this the fifth deepest X-ray field to date: the survey contains 256 sources detected down to a 0.5--2\,keV flux limit f$_{\rm 0.5-2}$=6$\times$10$^{-17}$\,\flu.

In this work, we present the spectroscopic and photometric redshift information for the \cha\ J1030 catalog \citep{nanni20}. The paper is organized as follows: in Section~\ref{sec:sample} we describe the \cha\ J1030 sample, its X-ray properties, and the optical/infrared counterparts identified in previous works. In Section~\ref{sec:spectra} we report the results of the spectroscopic campaigns aimed at increasing the redshift completeness of the sample, while in Section~\ref{sec:photo-z} we present the method we use to compute the X-ray sources photometric redshifts, and assess their quality using the spectroscopic redshifts as a reference. In Section~\ref{sec:catalog_description} we present the \cha\ J1030 redshift catalog that we make available to the public. In Section~\ref{sec:sample_properties} we present the main properties of the sample and discuss the presence of large-scale structures in the J1030 field, while in Section~\ref{sec:high-z} we focus on the {\it z}$\,>\,$3 subsample and report the \cha\ J1030 high--$z$ number counts. Finally, we summarize the results of this paper in Section~\ref{sec:summary}. 

Through the rest of the work, we assume a flat $\Lambda$CDM cosmology with H$_0$=69.6\,km\,s$^{-1}$\,Mpc$^{-1}$, $\Omega_m$=0.29 and $\Omega_\Lambda$=0.71 \citep{bennett14}. Errors are quoted at the 90\,\% confidence level, unless otherwise stated.

\section{The \cha\ J1030 field}\label{sec:sample}
\subsection{X-ray data}\label{sec:x-ray_cat}
\citet{nanni20} reported the results of a $\sim$479\,ks \cha\ observation over an area of 335\,arcmin$^2$ in the J1030 field. Such field is centered on the $z$=6.31 quasar SDSS J1030+0525, which is one of the most massive \citep[M$_{\rm BH}>$10$^9$\,M$_\odot$;][]{kurk07,derosa11} QSOs detected at {\it z}$\,>\,$6 by the Sloan Digital Sky Survey (SDSS), and the first one to reside in a spectroscopically confirmed overdensity at {\it z}$\,>\,$6 \citep{mignoli20}. 

The \cha\ catalog contains 256 sources, which are detected down to a 0.5--2\,keV flux limit $f_{\rm 0.5-2}$=6$\times$10$^{-17}$\,\flu. Such a deep flux limit makes J1030 an ideal region to search for intrinsically faint and/or heavily obscured AGN. Optical/nIR counterparts selected in the $r$, $z$, J and 4.5\,$\mu$m were associated with the X-ray sources using standard likelihood-ratio matching techniques. Out of 256 X-ray sources, seven are stars, while six lack optical/nIR counterpart. We note that the reliability of the \cha\ J1030 catalog is $\sim$96\,\%, which means that we expect to have $\sim$10--11 spurious sources in the catalog. As shown in previous X-ray surveys, these spurious detections are more likely to be found among the objects with no optical/nIR counterpart \citep[][]{luo17}. On the other hand, it is also possible that some of these objects are very high-redshift, faint AGN that are detected only in the X-rays. We will further explore this possibility in Section \ref{sec:high-z}. For our analysis, we will mostly focus on the 243 \cha\ J1030 extragalactic sources for which \citet{nanni20} reported a counterpart.

\begin{figure*}[htbp]
 \centering
\includegraphics[width=1.\linewidth]{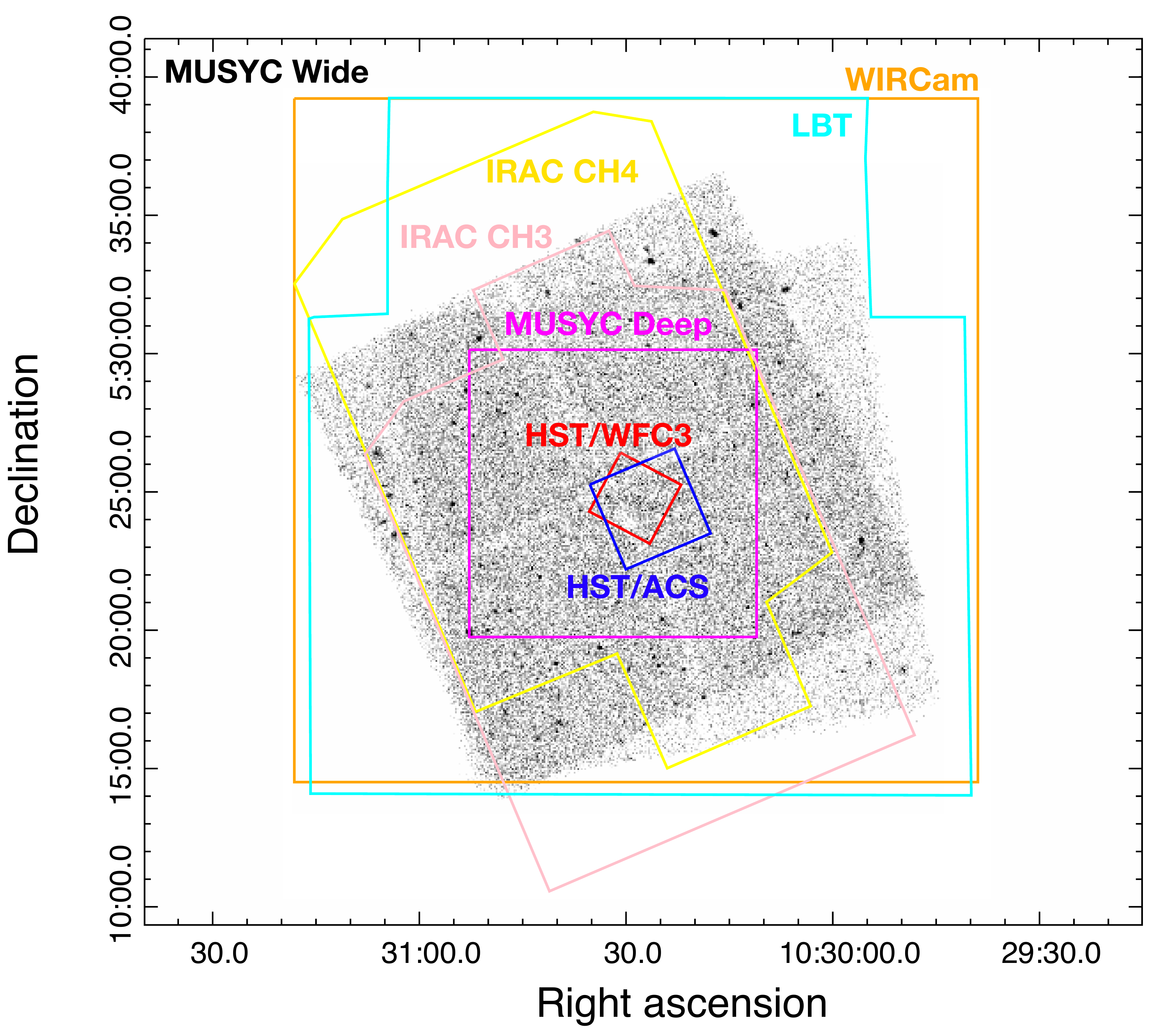}
\caption{\normalsize 
Smoothed 0.5--7\,keV \textit{Chandra} ACIS--I image of the SDSS J1030+0524 field. 
The regions covered by the photometric catalogs used in this work are also shown in different colors. The whole J1030 field has also been imaged in the mid-infrared IRAC channels 1 and 2.
}\label{fig:j1030_mosaic}
\end{figure*}

\subsection{Multiwavelength coverage}\label{sec:multiwave_data}
We report in Table \ref{tab:photometry_summary} the optical and nIR surveys from which we obtained the photometric information used to compute our photometric redshifts. The regions covered by these surveys are shown in Figure \ref{fig:j1030_mosaic} on top of the \cha\ J1030 0.5--7\,keV image.

The J1030 field was originally observed as part of the Multiwavelength Survey by Yale-Chile (MUSYC), and three photometric catalogs were made available. One (hereafter, MUSYC BVR) was obtained by detecting sources in a 30$^\prime\times$30$^\prime$, stacked BVR image, down to a 5\,$\sigma$ AB magnitudes $\sim$26 in the B and V bands \citep{gawiser06}, and contains photometric information in the U, B, V, R, $i$ and $z$ bands. The remaining two MUSYC catalogs were obtained by detecting sources in the K band: the K Deep catalog \citep{quadri07} covers the central 10$^\prime\times$10$^\prime$ area down to a 5\,$\sigma$ K$_{\rm AB}$=23, while the K Wide one \citep{blanc08} covers the same 30$^\prime\times$30$^\prime$ area analyzed in MUSYC BVR, and reaches a 5\,$\sigma$ magnitude K$_{\rm AB}$=21. Both K-band catalogs contain photometric information in the U, B, V, R, $i$, $z$ and K bands; the K Deep catalog also contains J and H magnitude information.  We include in our spectral energy distribution (SED) all the data points in the  U, B, V, R, $i$, $z$, J, H and K MUSYC bands.

In 2012, our group performed a deep photometric campaign with the Large Binocular Telescope using the Large Binocular Camera (LBT/LBC). This campaign covered a 23$^\prime\times$25$^\prime$ region in the $r$, $i$ and $z$ bands, reaching 5\,$\sigma$ AB magnitude limits $r_{\rm AB}$=27.5, $i_{\rm AB}$=26.5, $z_{\rm AB}$=25.2 \citep{morselli14}. In 2015, a 24$^\prime\times$24$^\prime$ area was covered in the Y and J bands using the WIRCam nIR mosaic imager mounted on the Canada France Hawaii Telescope (CFHT/WIRCam): the results of this campaign are reported in \citet{balmaverde17}. The 5\,$\sigma$ AB magnitude limits of this survey are $Y_{\rm AB}$=23.8 and $J_{\rm AB}$=23.75.

We used the LBC \hbox{$z$-band} image, which had the best seeing (0.64$^{\prime\prime}$ full width at half maximum, FWHM) to perform a morphological classification of our sources. In particular, we classified as ``point-like'' all those objects brighter than $z_{\rm AB}$=24 and with \texttt{CLASS\_STAR} \citep[a parameter derived using SExtractor and used to separate star-shaped and extended sources,][]{bertin96} greater then 0.9 \citep[see][]{morselli14}.

A small part of the field has also been targeted with the Hubble Space Telescope (HST). More in detail, a 3.3$^\prime\times$3.3$^\prime$ region has been observed with the Advanced Camera for Surveys (ACS) in the $i_{\rm F775W}$ and $z_{\rm F850LP}$ bands \citep{stiavelli05,kim09}, down to an AB magnitude limit of 27.5. An even smaller region of 2$^\prime\times$2$^\prime$ has been imaged with the Wide Field Camera 3 (WFC3) in the H$_{\rm F160W}$ band, down to an AB magnitude limit of 27.5 \citep[][T. Morishita et al. in prep.]{damato20}. Out of the 256 \cha\ J1030 sources, 21 have an HST/ACS counterpart and 12 out of these 21 also have an HST/WFC3 counterpart. All the \cha\ J1030 sources within the area covered by the two HST surveys have an HST counterpart.

Finally, we included in our SED analysis mid-infrared (mIR) information from the \textit{Spitzer} Infrared Array Camera (IRAC). The whole J1030 field has been imaged in the IRAC channels 1 and 2 at 3.6\,$\mu$m and 4.5\,$\mu$m, respectively, down to AB magnitude limits m$_{\rm AB,CH1}$=22.7 and m$_{\rm AB,CH2}$=22.4\footnote{We point out that the IRAC maps show strong sensitivity gradients, therefore these limits do not necessarily apply to the whole field.} \citep{annunziatella18}.  We also performed aperture photometry at the position of the X-ray sources to obtain mIR magnitude measurements for the whole \cha\ J1030 sample \citep{nanni20}. We then used IRSA archival data\footnote{\url{https://irsa.ipac.caltech.edu/}} to extend our mIR photometric coverage to the IRAC channels 3 and 4 (5.8 and 8\,$\mu$m, respectively) down to AB magnitude limits m$_{\rm AB,CH3}$=22.1 and m$_{\rm AB,CH4}$=21.8. These bands are particularly useful to better constrain high--redshift and/or heavily obscured sources.

\begingroup
\renewcommand*{\arraystretch}{1.5}
\begin{table*}
\centering
\scalebox{0.95}{
\vspace{.1cm}
 \begin{tabular}{ccccc}
 \hline
 \hline
 Catalog & Filters & Area & 5\,$\sigma$ Depth & Reference \\
     &     &   & m$_{\rm AB}$\\
  \hline    
  MUSYC BVR & UBVRI$z$ & 30$^\prime\times$30$^\prime$ & 26, 26, 26, 25.8, 24.7, 23.6  & \citet{gawiser06}  \\
  LBT/LBC & $riz$ & 23$^\prime\times$25$^\prime$ & 27.5, 26, 25 & \citet{morselli14}\\
  HST/ACS & F775W & 3.3$^\prime\times$3.3$^\prime$ & 27.5 & \citet{stiavelli05} \\
  HST/ACS & F850LP & 3.3$^\prime\times$3.3$^\prime$ & 27.5 & \citet{kim09} \\
  WIRCAM/CFHT & YJ & 24$^\prime\times$24$^\prime$ & 23.8, 23.75 & \citet{balmaverde17}\\
  HST/WFC3 & F160W & 2$^\prime\times$2$^\prime$ & 27.5 & \citet{damato20}\\
  MUSYC K Deep & UBVRI$z$JHK & 10$^\prime\times$10$^\prime$ & \small{25.6, 25.9, 25.9, 25.8, 25.8, 23.8, 23.4, 23.6, 23.2} & \citet{quadri07}\\
  MUSYC K Wide & UBVRI$z$K & 30$^\prime\times$30$^\prime$ & \small{25.6, 25.9, 25.9, 25.8, 25.8, 23.8, 21.9} & \citet{blanc08}\\
  \textit{Spitzer}/IRAC & CH1-2 & 35$^\prime\times$35$^\prime$ & 22.7, 22.4 & \citet{annunziatella18}\\
  \textit{Spitzer}/IRAC & CH3-4 & 22$^\prime\times$15$^\prime$ & 22.1, 21.8 & IRSA Archive\\
  \hline
	\hline
\end{tabular}}
	\caption{\normalsize Properties of the photometric catalogs used in this work.
	}
\label{tab:photometry_summary}
\end{table*}
\endgroup

\section{Spectroscopic coverage of the \cha\ J1030 field}\label{sec:spectra}
In the past few years, our group performed an extensive campaign to spectroscopically follow-up the X-ray selected J1030 sources, using several multi-object spectrographs: the Multi-Object Dual Spectrograph \citep[MODS,][]{pogge10} at the LBT, the FOcal Reducer and Spectrograph \citep[FORS2,][]{appenzeller98} and the Multi Unit Spectroscopic Explorer \citep[MUSE,][]{bacon10}, both mounted on the Very Large Telescopes (VLTs) at ESO, and, finally, the DEep Imaging Multi-Object Spectrograph (DEIMOS) on the 10m Keck II telescope \citep[][]{faber03}. Data reduction was performed with different software packages: standard IRAF procedures were used for the FORS2 and DEIMOS spectroscopic data, while the MODS data were reduced by the INAF--LBT Spectroscopic Reduction Center\footnote{\url{http://www.iasf-milano.inaf.it/software}} in Milan, where the LBT spectroscopic pipeline was developed \citep{garilli12}.

\subsection{The INAF--LBT Strategic Program}
\begin{figure}[htbp]
 \centering
\includegraphics[width=1.\linewidth]{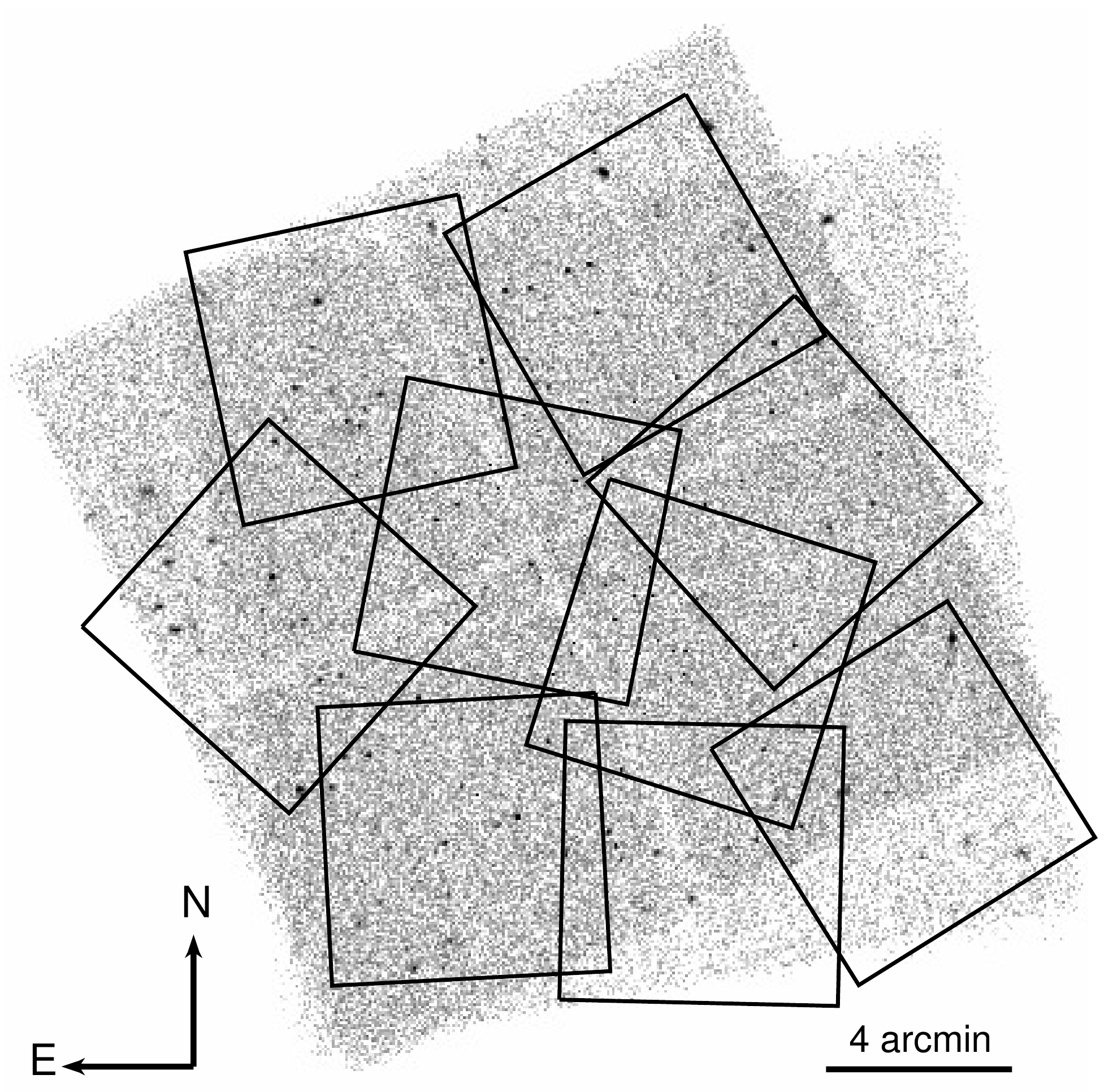}
\caption{\normalsize 
Smoothed 0.5--7\,keV \textit{Chandra} ACIS--I image of the SDSS J1030+0524 field. The position of the nine LBT/MODS masks used to obtain spectroscopic redshifts of a sample of \cha\ J1030 sources are shown with black boxes.
}\label{fig:field_w_MODS_masks}
\end{figure}

The vast majority ($\sim 95\%$) of the spectroscopic redshifts presented in this work were obtained through a 52 hours INAF--LBT Strategic Program that was granted to our group in 2017 (Program ID 2017/2018 \#18). The observations were performed between February 2018 and May 2019, and include 36 hours of multi-object optical spectroscopy with MODS (9 masks, 4 hours each), and 16 hours of long-slit near-IR spectroscopy with the LBT Utility Camera in the Infrared (LUCI; 4 slits, four hours each). The LBT is made of two telescope units of 8.4m diameter each that are equipped with similar instrumentation. In particular, nearly identical optical spectrographs, MODS1 and MODS2, and near-IR spectrographs, LUCI1 and LUCI2, are mounted on the two telescopes. When the LBT is used in binocular mode with the same instrument pair, the total night time on target is then halved. We observed most of the MODS masks and LUCI slits in binocular mode using the same instrument configuration on the two telescopes. The layout of the MODS masks (see Figure~\ref{fig:field_w_MODS_masks}) was designed to include as many as possible X-ray source counterparts in the slits, avoiding at the same time overcrowding that could compromise the spectral extraction and analysis process. Each mask includes a variable number (from 14 to 24 slits) of X-ray counterparts depending on the geometry: a total of 159 optical counterparts of X-ray sources were included in the slits. When possible, non-X-ray targets were also included in the masks: we will further mention these targets in Section~\ref{sec:structures}.
MODS spectra were obtained using both the blue G400L (3200$-$5900~\AA) and the red G670L (5400$-$10000~\AA) gratings on the blue and red channels in dichroic mode. The wide spectral range provided by the adopted configuration has proved extremely useful in determining the redshift and spectral classification of the observed objects, despite the instrument sensitivity drops in the overlapping spectral region of the dichroic. MODS observations allowed us to obtain a redshift measurement for 107 out of 159 observed X-ray sources (including two Galactic stars); the remaining 52 sources have low signal-to-noise ratio and do not show any clear feature that can be identified. The efficiency in the redshift measurement is almost 100\% up to magnitude $r_{\rm AB}$=23.5, and then progressively decreases. Only a few targets with extremely intense emission lines are indeed spectroscopically identified at magnitudes $r_{\rm AB}\gtrsim\,$25.

 A few more redshifts were obtained trough near-IR long-slit spectroscopy with LUCI. We adopted 4-arcmin-long slits and choose their position angle so as to include two X-ray source counterparts per slit. LUCI targets were primarily selected to be faint in the optical ($r_{\rm AB}>$25) but relatively bright in the near-IR ($J<23$), i.e. they were candidate obscured, high-$z$ sources beyond reach for MODS but promising for near-IR spectroscopy. We targeted seven X-ray red counterparts (plus one radio source) in four LUCI pointings, and measured the redshift for three of them, through the detection of the H$\alpha$+[N\,\textsc{ii}] emission lines complex. One of these three objects, XID 189, is indeed a heavily obscured, type-2 AGN at z=1.7 at the center of the protocluster described in \citet{gilli19}.

\subsection{Other follow-up programs}
Additional redshifts were acquired as part of spectroscopic follow-ups primarily aimed at confirming the galaxy overdensity around the z=6.31 J1030 QSO \citep{morselli14,balmaverde17}.  In particular we used spare slits during  follow-ups of color-selected {\it z}$\,\sim\,$6 candidates  with the FORS2/VLT and DEIMOS/Keck multi-object spectrographs. Moreover, the central 1'$\times$1' region of the J1030 field was observed in 2016 by MUSE for a total of 6.4~hours of exposure time. We retrieved and analyzed the archival observation \citep{gilli19} and obtained redshifts for the three X-ray sources in the MUSE field. All these observations, allowed us to spectroscopically identify seven further X-ray sources. Finally, the counterparts of ten X-ray sources (excluding the $z$=6.31 quasar SDSS~J1030+0524 itself) had already been  observed as part of the SDSS \citep[][]{ahumada20}. In total, 174 out of 243 extragalactic X-ray sources with an optical/nIR counterpart have been targeted at least once in our spectroscopic campaigns or had archival SDSS spectra. 

\subsection{Redshift measurements and spectral classification of the X-ray sources counterparts}\label{sec:spec-z_class}

We determined a spectroscopic redshift and classification for 123 out of the 174 \cha\ J1030 sources for which we have an available spectrum. The redshifts measurement was done by means of automatic software (IRAF: {\tt rvidlines} and {\tt xcsao}) and checked through visual inspection of the 1D and 2D spectra. The spectral analysis was performed using the IRAF interactive package {\tt splot} and, according to the measured features, the X-ray source counterparts were grouped into four classes: $i$) Broad-Line AGN (BL-AGNs), i.e., objects with spectra showing broad (FWHM$>$1500\,km\,s$^{-1}$) emission lines; $ii$) Narrow Line AGN (NL-AGNs), i.e., objects with spectra showing narrow (FWHM$<$1500\,km\,s$^{-1}$) high ionization emission lines\footnote{E.g., [OIII]$\lambda$5007; [NeV]$\lambda$3427; HeII$\lambda$1640; and CIV$\lambda$1549.}; $iii$) Emission Line Galaxies (ELGs), i.e., sources whose spectra do not show any detectable AGN features, but a flat or bluish continuum and low--ionization emission lines that are compatible with a typical star formation activity; $iv$) Early-Type Galaxies (ETGs), i.e., sources whose spectra are dominated by a red continuum with pronounced absorption lines, suggestive of an evolved stellar population, with absent or very faint emission lines. We stress that the spectral classification does not always describe the source nature, but it simply reflects the presence of spectral features that may, or may not, be traced back to the presence of an AGN. In particular, the NL-AGN classification is often connected to the presence, or to the detectability, in the observed spectral range of high ionization emission lines which can be directly linked to nuclear activity. It is worth noting that more than half of the sources classified as ELGs lie at z$\sim$1, where the brightest high ionization lines are out of the optical range.

\begin{figure*} 
\begin{minipage}[b]{.49\textwidth} 
 \centering 
 \includegraphics[width=1.0\textwidth]{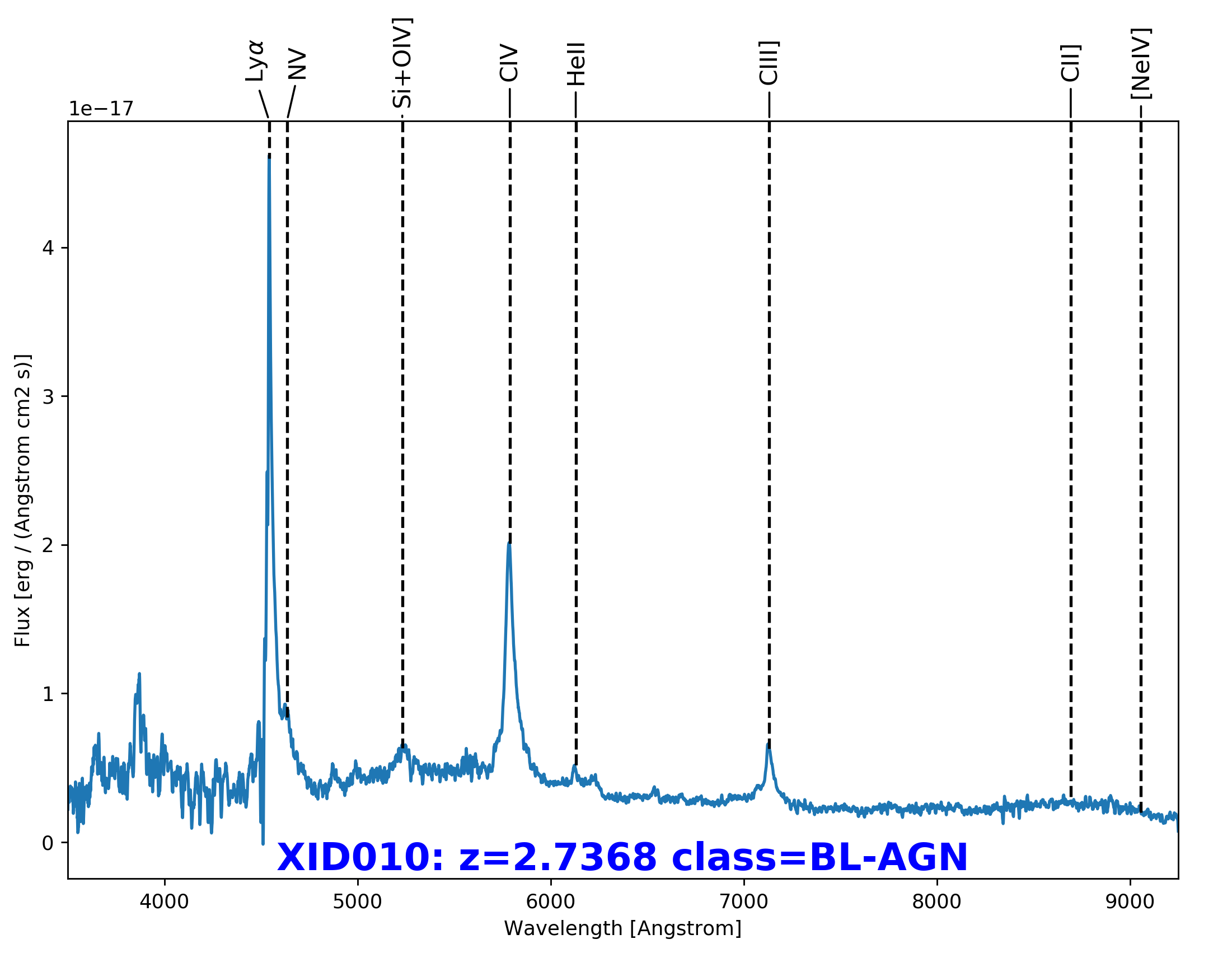} 
 \end{minipage} 
\begin{minipage}[b]{.49\textwidth} 
 \centering 
 \includegraphics[width=1.0\textwidth]{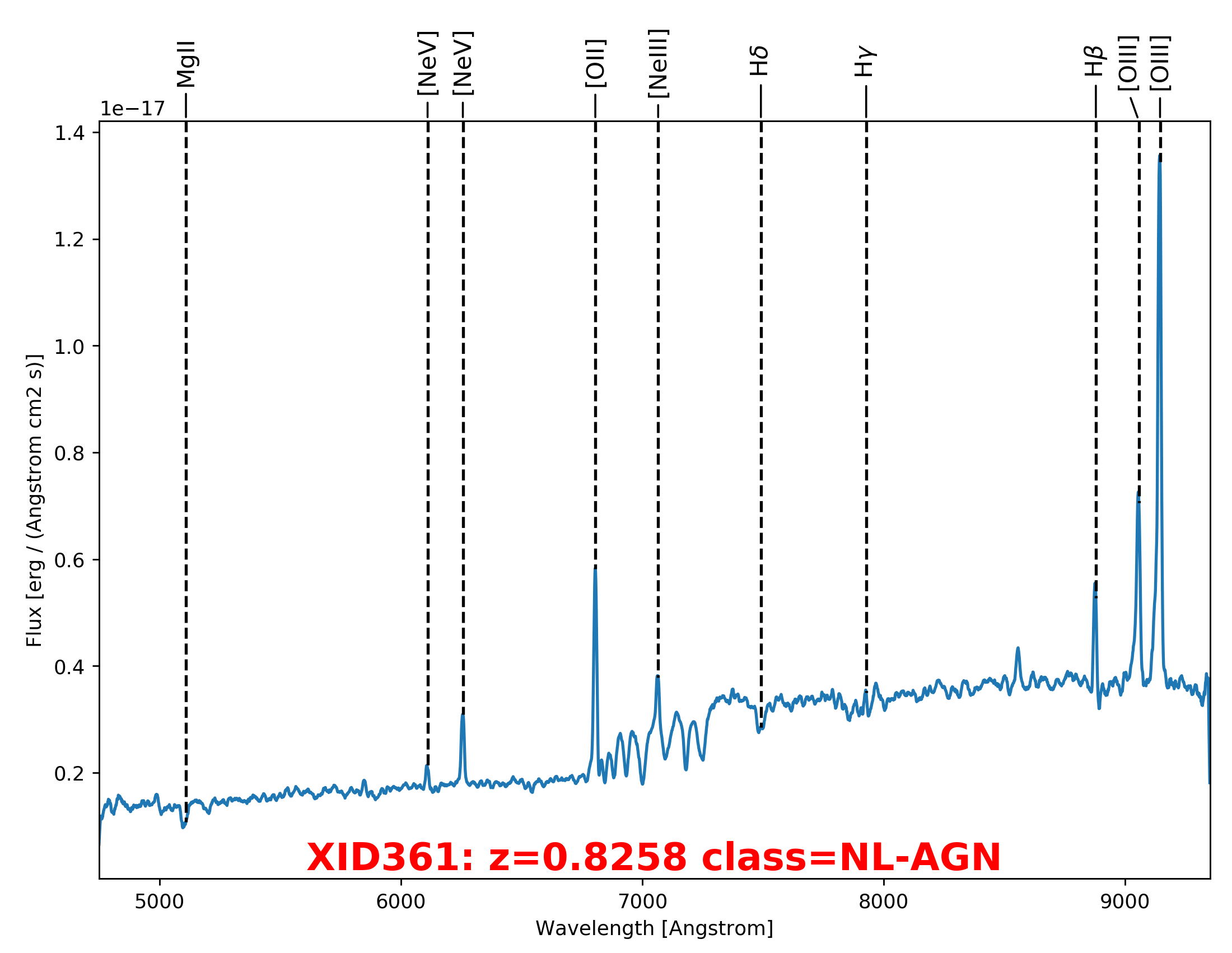} 
 \end{minipage} 
\begin{minipage}[b]{.49\textwidth} 
 \centering 
 \includegraphics[width=1.0\textwidth]{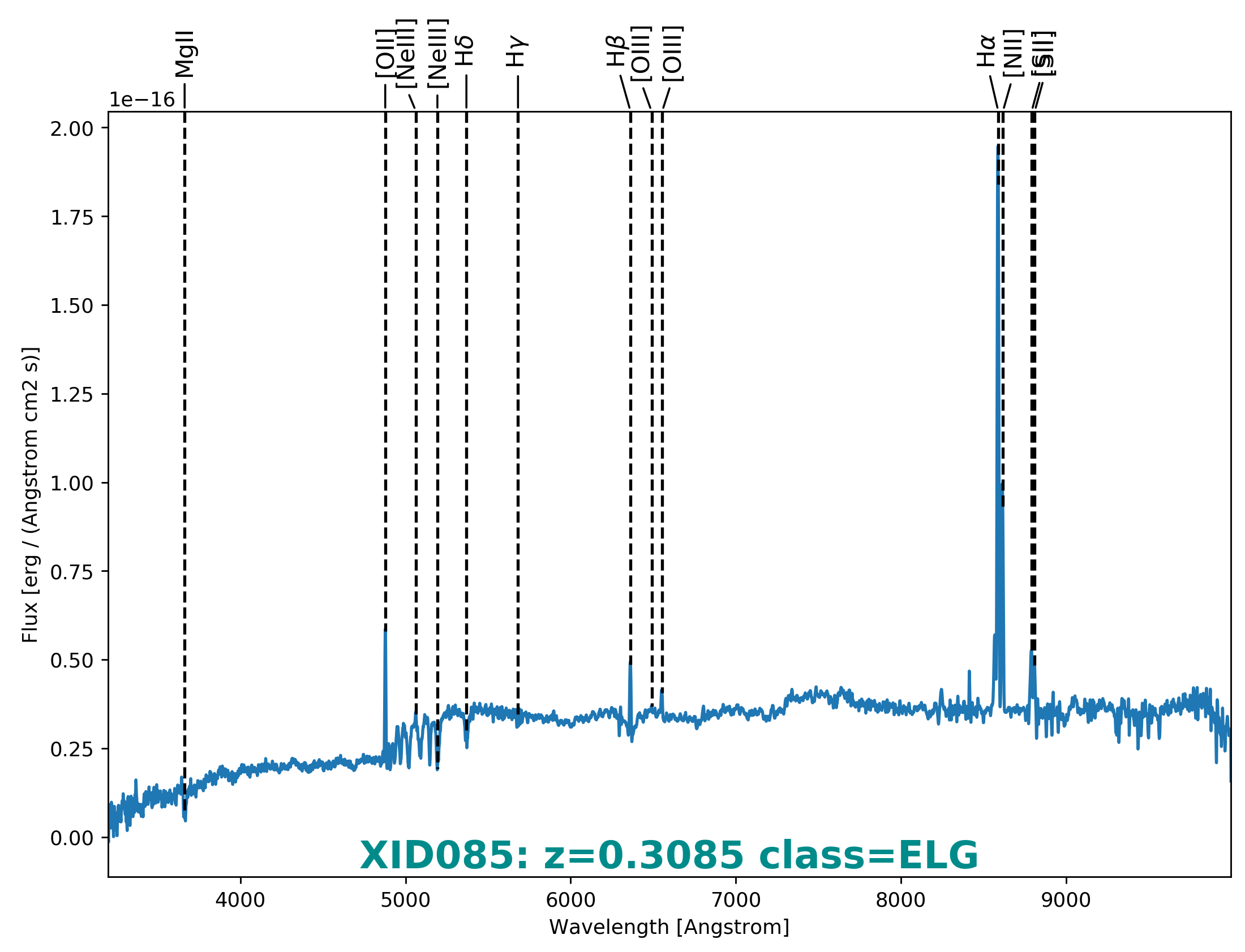} 
 \end{minipage} 
\begin{minipage}[b]{.49\textwidth} 
 \centering 
 \includegraphics[width=1.0\textwidth]{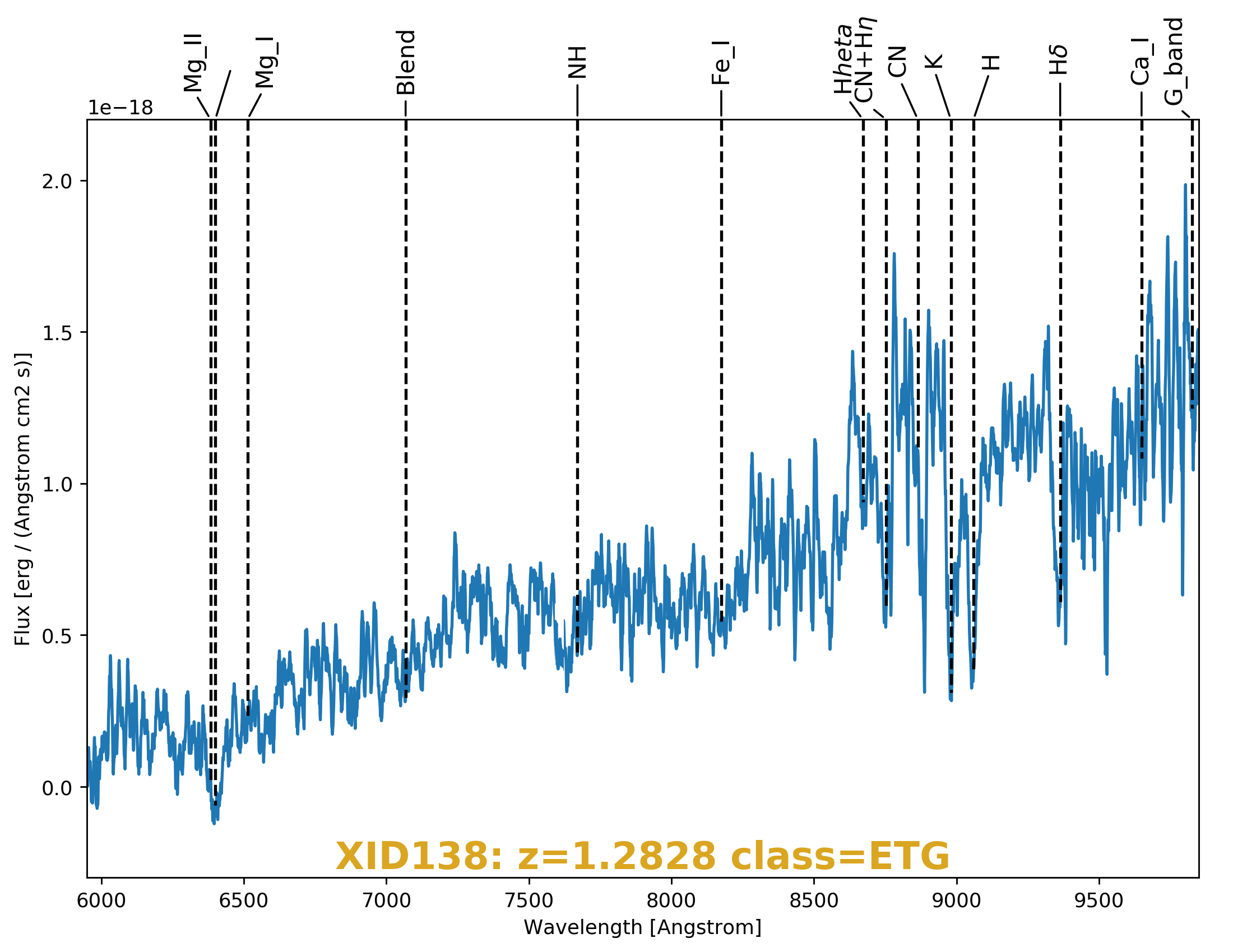} 
 \end{minipage} 
\caption{\normalsize 
Example of MODS spectra of four \cha\ J1030 optical counterparts: all the spectra have been obtained with the LBT strategic program. From the top left, moving clockwise: XID-10 (broad-line AGN); XID-361 (narrow-line AGN); XID-138 (early type galaxy); XID-85 (emission-line galaxy).
}\label{fig:spectra_example}
\end{figure*}

We report in Table \ref{tab:spec_type} the number of sources for each spectral class: 63 sources are classified as AGN, either broad-line (43 objects) or narrow-line (20 sources). Other 60 sources have spectra dominated by the host galaxy emission: 28 are emission-line galaxies (ELGs), while 32 are early-type galaxies (ETGs). Among the ETGs, a majority of objects (21, 66\,\%) also show faint emission lines perhaps indicative of obscured and/or low-efficiency SMBH accretion. We report an example of a spectrum for each class of sources in Figure \ref{fig:spectra_example}. All spectra are available online at \url{http://j1030-field.oas.inaf.it/xray_redshift_J1030.html}.

\begin{figure}[htbp]
 \centering
\includegraphics[width=1.\linewidth]{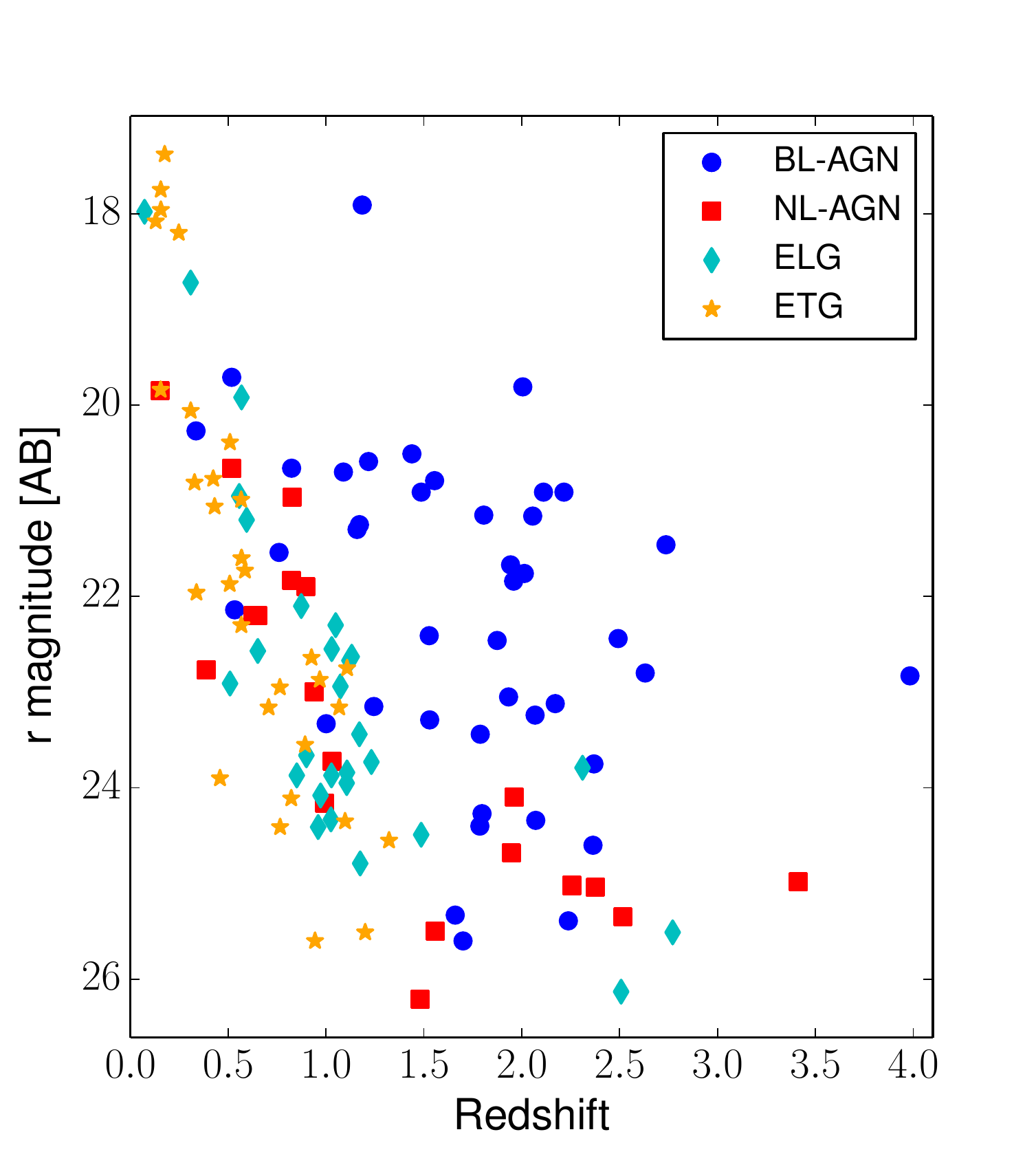}
\caption{\normalsize 
Total AB magnitude in the $r$ band from \citet{nanni20} as a function of spectroscopic redshift for 122 \cha\ J1030 sources with $z_{\rm spec}$: we do not include in this plot the $z$=6.3 QSO SDSSJ1030+0525, which is not detected in the $r$-band. BL-AGNs are plotted as blue circles, NL-AGN as red squares, ELGs as cyan diamonds, and ETGs as orange stars. 
}\label{fig:z_vs_r-mag_w_spec_type}
\end{figure}

In Figure~\ref{fig:z_vs_r-mag_w_spec_type} we plot the $r_{\rm AB}$~magnitude against the spectroscopic redshift for the 123 identified \cha \ sources, with different symbols according to their spectral class: as expected, the unobscured (Broad-Line) AGNs stand out as the most luminous objects at all redshifts, whereas obscured (Narrow-Line) AGNs occupy the same loci as host--galaxy dominated sources.

We will use the spectroscopic information to assess the reliability of our photometric redshifts in Section \ref{sec:spec_vs_phot}.

\begingroup
\renewcommand*{\arraystretch}{1.5}
\begin{table}
\centering
\scalebox{1.}{
\vspace{.1cm}
 \begin{tabular}{ccc}
 \hline
 \hline
 Class & N$_{\rm src}$ & Agree \\
  \hline    
  BL-AGN & 43 & 28 (65\,\%) \\
  NL-AGN & 20 & 14 (70\,\%) \\
  ELG & 28 & 25 (89\,\%) \\
  ETG (with hybrids) & 32 (31) & 30 (97\,\%) \\
  \hline
  Overall & 123 (122) & 97 (80\,\%)\\
  \hline
	\hline
\end{tabular}}
	\caption{\normalsize Number of sources by spectral classification, and number of objects for spectral class where the spectroscopic and photometric redshifts are in agreement, i.e., where $\Delta z$=||$z_{\rm phot}$-$z_{\rm spec}$||$<$0.15(1+$z_{\rm spec}$). BL-AGNs are broad-line active galactic nuclei, NL-AGNs are narrow-line active galactic nuclei, ELGs are emission-line galaxies, and ETGs are early-type galaxies (including the objects with faint emission line detected in the spectrum). We note that XID 143 is an ETG located nearby a bright source for which no reliable photometry (and consequently no photometric redshift) could be estimated.
	}
\label{tab:spec_type}
\end{table}
\endgroup

\section{Photometric redshift of the \cha\ J1030 sources}\label{sec:photo-z}
\subsection{Spectral energy distribution fitting procedure}\label{sec:SED_fitting}

\begin{figure}[htbp]
 \centering
\includegraphics[width=1.\linewidth]{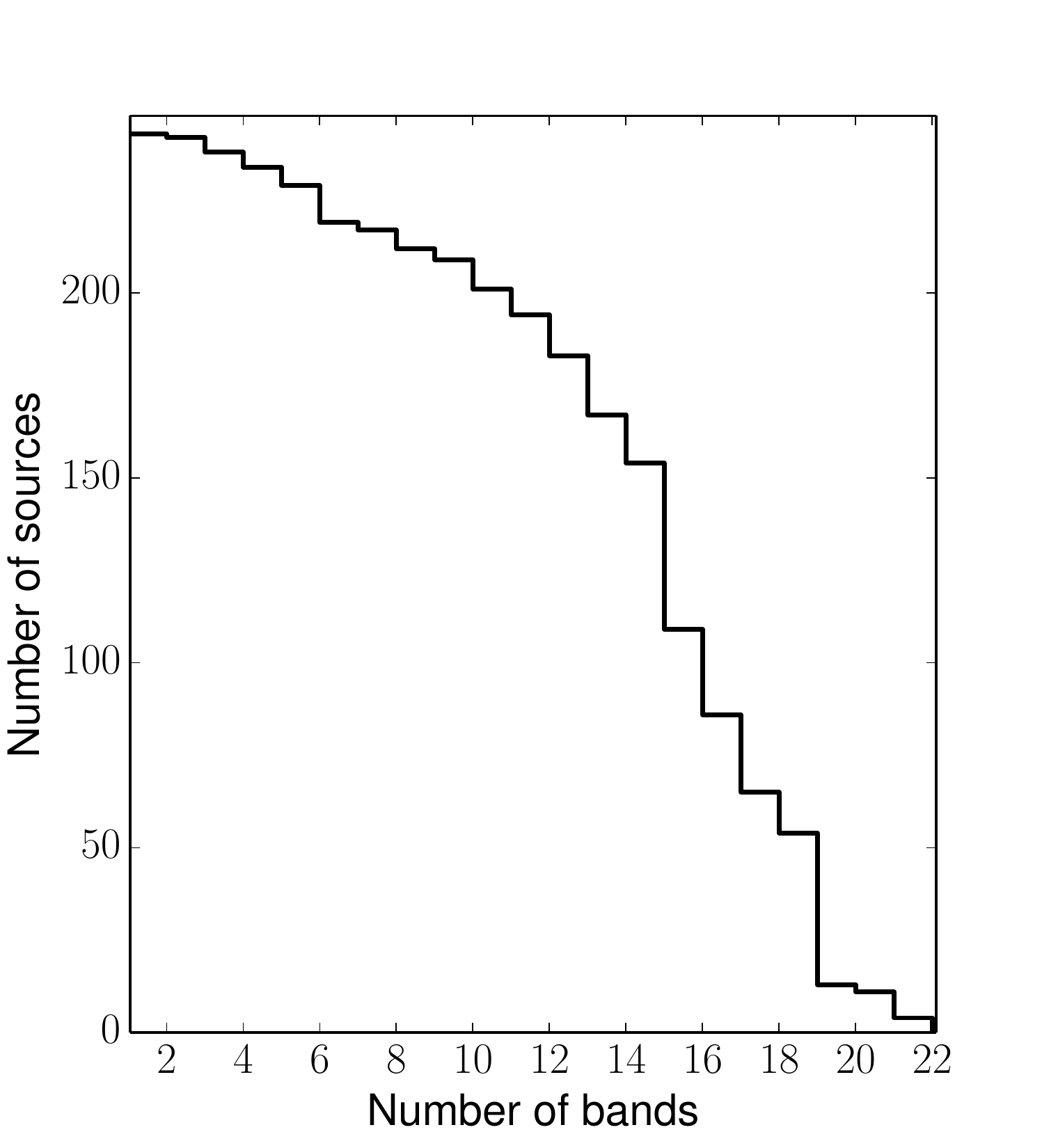}
\caption{\normalsize 
Cumulative distribution function of the number of bands in which a source has been detected.
}\label{fig:histo_nband}
\end{figure}

In Section \ref{sec:multiwave_data}, we presented the multi-year effort which allowed us to image the J1030 field with a large variety of facilities in the optical and near infrared. Overall, the \cha\ J1030 field was observed in up to 21 different bands: while only four objects were detected in all the 21 bands, 154 out of 243 objects (63\,\% of the sources with a photometric redshift) have a detection in at least 14 bands. We report in Figure \ref{fig:histo_nband} the cumulative distribution function of the number of bands in which a source has been detected. 

We use the \texttt{Hyperz} code \citep{bolzonella00} to fit our SEDs. The code fits a range of SED models to the data-points and determines the best-fit parameters through a standard minimum $\chi^2$ approach. More in detail, the best fit solution is the one where 

\begin{equation}
    \chi^2(z)=\sum_{i=1}^{N_{\rm data}}\frac{(F_{\rm obs,i}-k\times F_{\rm temp,i}(z))^2}{\sigma_i^2}
\end{equation}

is minimized. The redshift value varies within the range $z$=[0--7], with steps $\Delta z$=0.02. N$_{\rm data}$ is the number of SED data-points, F$_{obs,i}$ is the observed flux for a given data-point in a certain band, F$_{temp,i}$($z$) is the expected model flux in the same band, $\sigma_i$ is the observed flux uncertainty, and $k$ is a normalization constant. If a source is not detected in a given band, we assume F$_{\rm obs,i}$=$\sigma_{\rm i}$=$m_{\rm UL,i}$/2, where $m_{\rm UL,i}$ is the upper limit on the magnitude in the $i$-th band. 

The SED fitting procedure is performed twice for each object; in all cases, we adopt the \citet{calzetti00} extinction law to take reddening into account.
First we use 75 templates dominated by stellar emission, following the work by \citet{ilbert13}. More in detail, 19 out of 75 are empirical templates from the SWIRE template library \citep{polletta07}:  7 for elliptical galaxies, 12 for different classes of spiral galaxies (S0, Sa, Sb, Sc, Sd and Sdm). Other 12 templates describe the SED of starburst galaxies, although no emission lines are included in the template. Finally, the remaining  44 templates are based on the \citet{bruzual03} stellar population synthesis models, and have been first introduced in \citet{ilbert09,ilbert13}. We allowed the absolute, non-absorption corrected magnitude of the sources to vary in the range $r_{\rm abs, AB}$=[--30:--16]: we note that this range is fairly conservative, particularly at the high-luminosity end. For each source, we therefore obtain a ``galaxy'' best-fit photometric redshift, $z_{phot,gal}$ with an associated $\chi^2_{\rm gal}$ (i.e., the minimum $\chi^2$($z$) value among those computed by means of Equation~1).

We then fit our SEDs using the 30 templates originally reported in \citet[][see their Table 2 for a complete description]{salvato09}. In most of these templates, the AGN contribution is taken into account: in some cases the AGN is fully dominant,  while in many other the template is a hybrid where both the AGN and the host contribute to the overall SED, with the relative contribution of the two components varying in the range 10--90\,\%. These fractions are computed between 5000 and 5200\,\AA\ for templates of unobscured, Type 1 AGN, and between 0.9 and 1\,$\mu m$ for templates of obscured, Type 2 AGN. As it can be seen from the SEDs plotted in Figure~3 of \citet{salvato09}, the ratios computed in the above-mentioned bandwidths are a good approximation of the bolometric ones.

We allowed the absolute, non-absorption corrected magnitude of the sources to vary in the range $r_{\rm abs}$=[--30:--21]. We choose a narrower magnitude range with respect to the one we used for the \citet{ilbert09} templates, since when comparing our photometric redshifts with the spectroscopic ones (see Section \ref{sec:spec_vs_phot}) we find that allowing for fainter luminosities can lead to several erroneous low--$z$ solutions, particularly for those sources where the SED is AGN--dominated, with implausibly low AGN luminosities. By means of the quantity $\chi^2_{\rm Hyb}$, defined as in Equation~1, we compute a hybrid (hereafter ``Hyb'') best-fit photometric redshift, $z_{phot,Hyb}$.

\begin{figure}[htbp]
 \centering
\includegraphics[width=1.\linewidth]{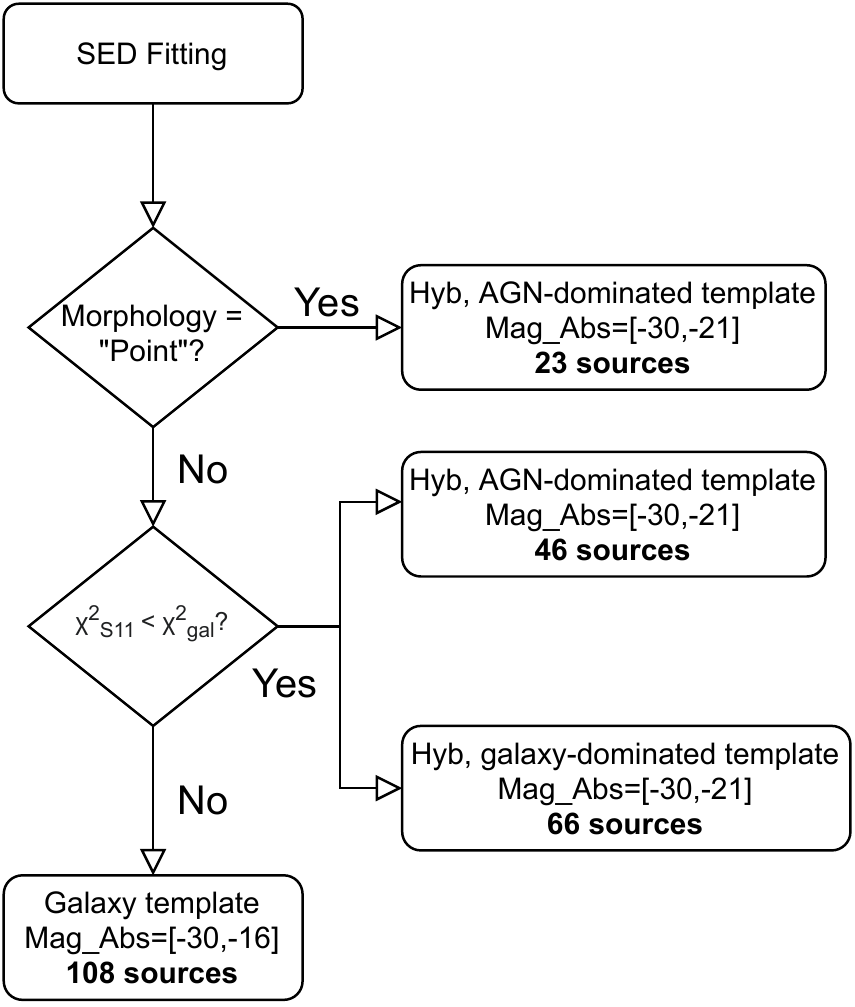}
\caption{\normalsize 
Flow chart of the procedure we adopted to choose the best-fit SED template.
}\label{fig:flow_template}
\end{figure}

We then use the flow chart reported in Figure \ref{fig:flow_template} to determine which of the two photometric redshifts should be chosen as best-fit $z_{\rm phot,best}$. The first criterion we adopt is based on the optical morphology of the source, since it has been shown that using the morphological information can help in significantly improve the photometric redshifts reliability \citep[e.g.,][]{salvato09,salvato11,ananna17}. More in detail, following the approach discussed in \citet{ananna17}, if an object is classified as point-like we choose a Hyb AGN--dominated template solution (i.e., one where the AGN contribution is greater or equal to 50\,\%)  even in those cases where the galaxy $\chi^2$ is smaller than the Hyb one. We then assign to these objects an ``AGN'' SED. Based on the STAR\_CLASS parameter we mentioned earlier in the text, there are 23 point-like sources among the 243 \cha\ J1030 extragalactic objects, and 9 of them have $\chi^2_{\rm Hyb}>\chi^2_{\rm gal}$ (which means that the ``galaxy'' solution would have been statistically favored). 

For the remaining 220 objects which are morphologically classified as ``extended'', we compare the two best-fit $\chi^2$ values: the number of degrees of freedom is the same for both families of templates, being d.o.f.=N$_{\rm filt}$-1, where N$_{\rm filt}$ is the number of photometric data points in the SED. Therefore, using $\chi^2$ is fully equivalent to using the reduced $\chi^2$. For the 108 sources having  $\chi^2_{\rm gal}<\chi^2_{\rm Hyb}$, we take $z_{\rm phot,best}$=$z_{phot,gal}$, and the SED of these objects is classified as ``galaxy''. Finally, there are 112 sources where $\chi^2_{\rm Hyb}<\chi^2_{\rm gal}$: here we take $z_{\rm phot,best}$=$z_{phot,Hyb}$. Among these 112 objects, we assign an ``AGN'' SED to the 46 sources whose SED is fitted with a template where the AGN contribution to the overall emission is greater or equal to 50\,\%. The remaining 66 sources, where the AGN contribution is subdominant, are associated with a ``galaxy'' SED.

In summary, 69 out of 243 sources (i.e., 28\,\% of the \cha\ J1030 extragalactic sample) have an AGN-dominated SED. Such a fraction is in close agreement with those measured in other deep X-ray surveys, such as, e.g., COSMOS \citep[33\,\%][; this fraction was obtained taking into account both the spectroscopic information, when available, and the best-fit SED template one]{marchesi16a}. We remark that this classification is purely based on the optical/near infrared SED fitting: based on their X-ray emission, we expect almost the totality of the \cha\ J1030 sources to host an AGN.

For all sources, we also computed a redshift probability distribution function (PDZ) which allows one to reliably estimate how well constrained are the photometric redshifts. The PDZs have a bin $\Delta z$=0.02, and we computed them following the approach presented in \citet[][see also, e.g., \citealt{ilbert09}]{vito18}. For each redshift bin, is PDZ($z$)$\propto e^{-\chi^2(z)/2}$, where $\chi^2$ is the $\chi^2$ value at redshift $z$ of the SED fitting computed by \texttt{Hyperz}. The PDZ($z$) values obtained this way are then are normalized to unit integral in the considered redshift range. For each source, we assign a PDZ value only to those redshifts whose absolute magnitude is within the permitted range we use in the SED fitting (i.e., $r_{\rm abs}$=[--30:--21] for sources best fitted with an AGN template, and $r_{\rm abs}$=[--30:--16] for sources best fitted with a galaxy template).

It has been shown \citep[e.g.,][]{yang14} that the PDZs derived directly from $\chi^2$ are usually too narrow, possibly due to a systematic underestimation of the photometric uncertainties. To quantify this effect, we computed the fraction of \cha\ J1030 sources where the spectroscopic redshift falls within the 68\,\%  confidence region of the photometric redshift  (i.e., within the narrowest redshift interval for which the integrated PDZ is equal to 0.68). For the PDZ to properly represent the photometric redshifts uncertainties, this fraction should be exactly 68\,\%: however, we find that in our sample the actual fraction is just 18\,\% (22 out of 122 objects with $z_{\rm spec}$).

To fix this discrepancy, we therefore searched for a constant $a$ that multiplies the uncertainties on the photometric measurements, so that $\sigma_{\rm corr}$=$a\sigma$, and, consequently, PDZ($z$)$_{\rm corr}\propto$ e$^{-\chi^2_{\rm corr}(z)/2}$, where 
\begin{equation}
\chi^2_{\rm corr}(z)=\sum_{i=1}^{N_{\rm data}}{\frac{(F_{\rm obs,i}-k\times F_{\rm temp, i}(z))^2}{a^2\sigma^2_i}}. 
\end{equation}

We note that the correction factor $a$ is the same for all redshifts, and it therefore does not affect the redshift best-fit value.

We find that we need a correction factor $a$=3.7 to have 68\,\% of our sources with a spectroscopic redshift ending up in the photometric redshifts 68\,\% confidence region. As a reference, \citet{vito18} computed $a$=2.1, 2.1 and 1.2 for the photometric redshifts measured in the CANDELS/\cha\ Deep Field-South by \citet{hsu14}, \citet{skelton14} and \citet{straatman16}, respectively. We note that all these works made use of narrow- and/or intermediate-band photometry that significantly improve the photometric redshifts accuracy, as we will further discuss in Section\,\ref{sec:compare_other_works}. Furthermore, we note that the $a$ value we obtain is driven by the presence in our spectroscopic sample of a large population of BL-AGNs  (43 out of 123, 35\,\%), a population of sources for which photometric redshifts are on average more complex to compute (see Section~\ref{sec:spec_vs_phot}). Indeed, if we compute $a$ for the non BL-AGN population we obtain $a$=2.8, closer to the values obtained in the CANDELS field.

In the rest of the work, and particularly when analyzing the \cha\ J1030 high-redshift sample (Section~\ref{sec:high-z}), we will use the corrected redshift probability distribution function, PDZ($z$)$_{\rm corr}$ when using the photometric redshifts. As for the spectra, all the SEDs and PDZs are made available online at \url{http://j1030-field.oas.inaf.it/xray_redshift_J1030.html}. SEDs and PDZs are available both in png and in ASCII format. 

\subsection{Comparison with spectroscopic redshifts}\label{sec:spec_vs_phot}
As mentioned in Section \ref{sec:spectra}, our spectroscopic follow-up campaign allowed us to obtain a spectroscopic redshift and a spectroscopic classification for 123 out of 243 \cha\ J1030 extragalactic sources.

We find that 97 out of 122 sources with $z_{\rm spec}$\footnote{We do not include in our comparison XID--143 ($z_{\rm spec}$=0.3282), since its photometry is contaminated by a much brighter nearby object, which prevented us from obtaining a reliable photometric redshift} have their best-fit photometric redshift in agreement with the spectroscopic one: we assume that the two redshifts are in agreement if ||$z_{\rm phot}$-$z_{\rm spec}$||/(1+$z_{\rm spec}$)$<$0.15 \citep[see, e.g.,][]{marchesi16a}. To further quantify the reliability of our photometric redshifts and their agreement with the spectroscopic ones, we compute the normalized median absolute deviation $\sigma_{\rm NMAD}$=1.48$\times$median(||$z_{\rm phot}$-$z_{\rm spec}$||/(1+$z_{\rm spec}$)) \citep{hoaglin83}. The whole sample of 122 sources with spectroscopic redshift has $\sigma_{\rm NMAD,All}$=0.065; as mentioned above, the fraction of outliers is $\eta_{\rm All}$=1-(97/122)=20.5\,\%.

Since this outlier fraction is somewhat higher than those measured in other works (see Section \ref{sec:compare_other_works}), we break down our spectroscopic sample in different subsamples, with the goal to determine which type of objects are more likely to have a wrong photometric redshift. We find two main indicators of a large outlier probability.

\begin{enumerate}
  \item Out of 43 spectroscopically classified broad-line AGN, there are 16 photometric outliers ($\eta_{\rm BL}$=35\,\%): this means that 65\,\% of the outliers in our sample are BL-AGN. This result is not unexpected, since the SED of luminous, unobscured AGNs are known to be more difficult to model. This is particularly true for those SEDs, like the ones analyzed in this work, that are based on broadband photometry only \citep[e.g.,][]{salvato11,ananna17}.
  \item Out of 20 morphologically classified point-like sources having a spectroscopic redshift (three point-like sources only have a photometric redshift), there are 8 photometric outliers ($\eta_{\rm Point}$=40\,\%). As for BL-AGNs, photometric redshifts for point-like sources are known to be more difficult to compute than those of extended objects, and indeed 18 out of 20 point-like sources are spectroscopically classified as BL-AGN. This high fraction of outliers in point-like sources is due to the fact that they are, for the most part, objects where the emission is AGN-dominated and SED templates are less accurate, especially when fitting SEDs lacking narrow-band photometry information. We also note that the morphological criterion described in Section \ref{sec:SED_fitting} allowed us to recover three photometric redshifts that would have otherwise been outliers based on the $\chi^2_{\rm gal}<\chi^2_{\rm AGN}$ classification alone.
\end{enumerate}

\begin{figure*} 
\begin{minipage}[b]{.45\textwidth} 
 \centering 
 \includegraphics[width=1.0\textwidth]{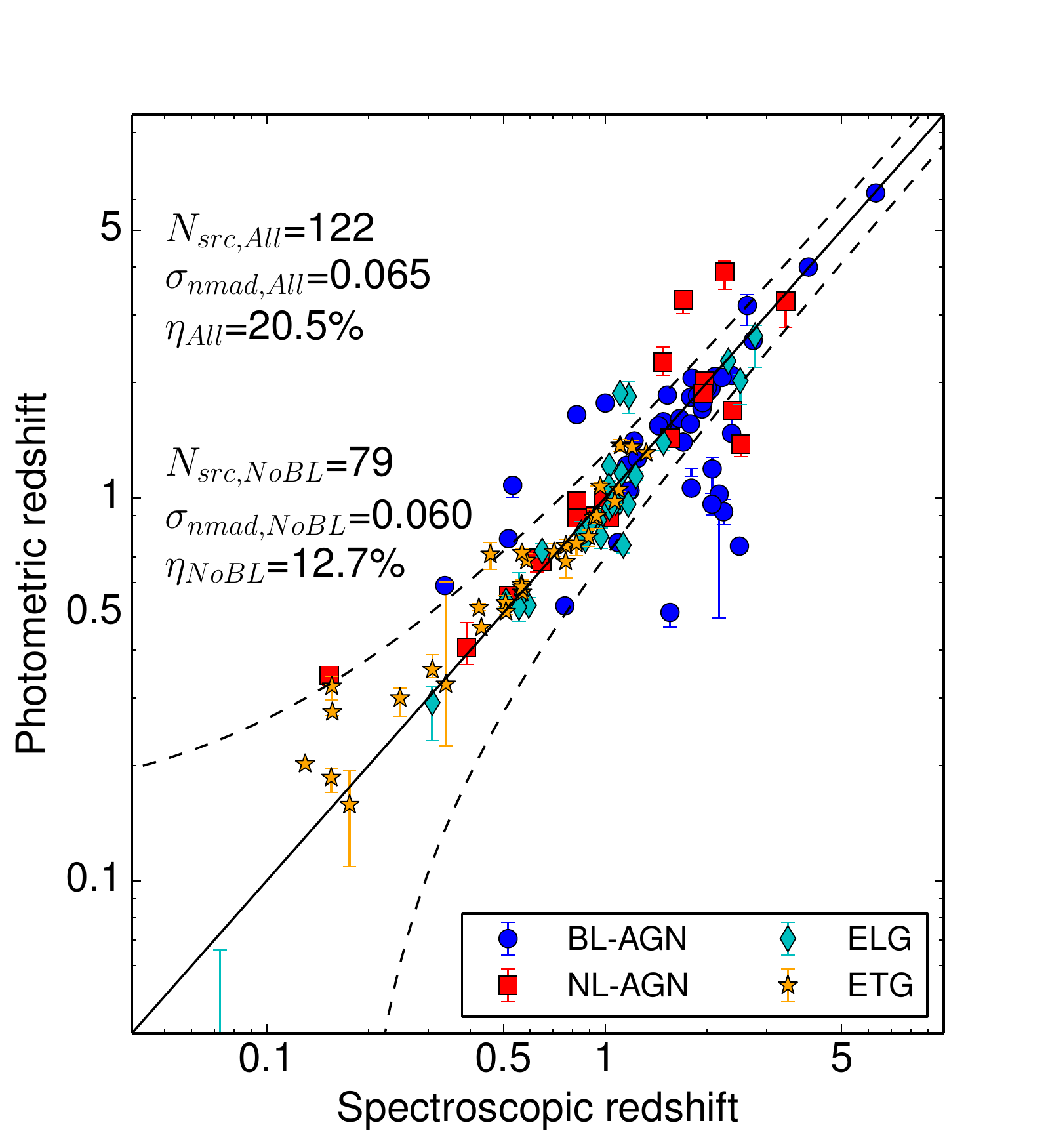} 
 \end{minipage} 
 \begin{minipage}[b]{.44\textwidth} 
 \centering 
 \includegraphics[width=1.0\textwidth]{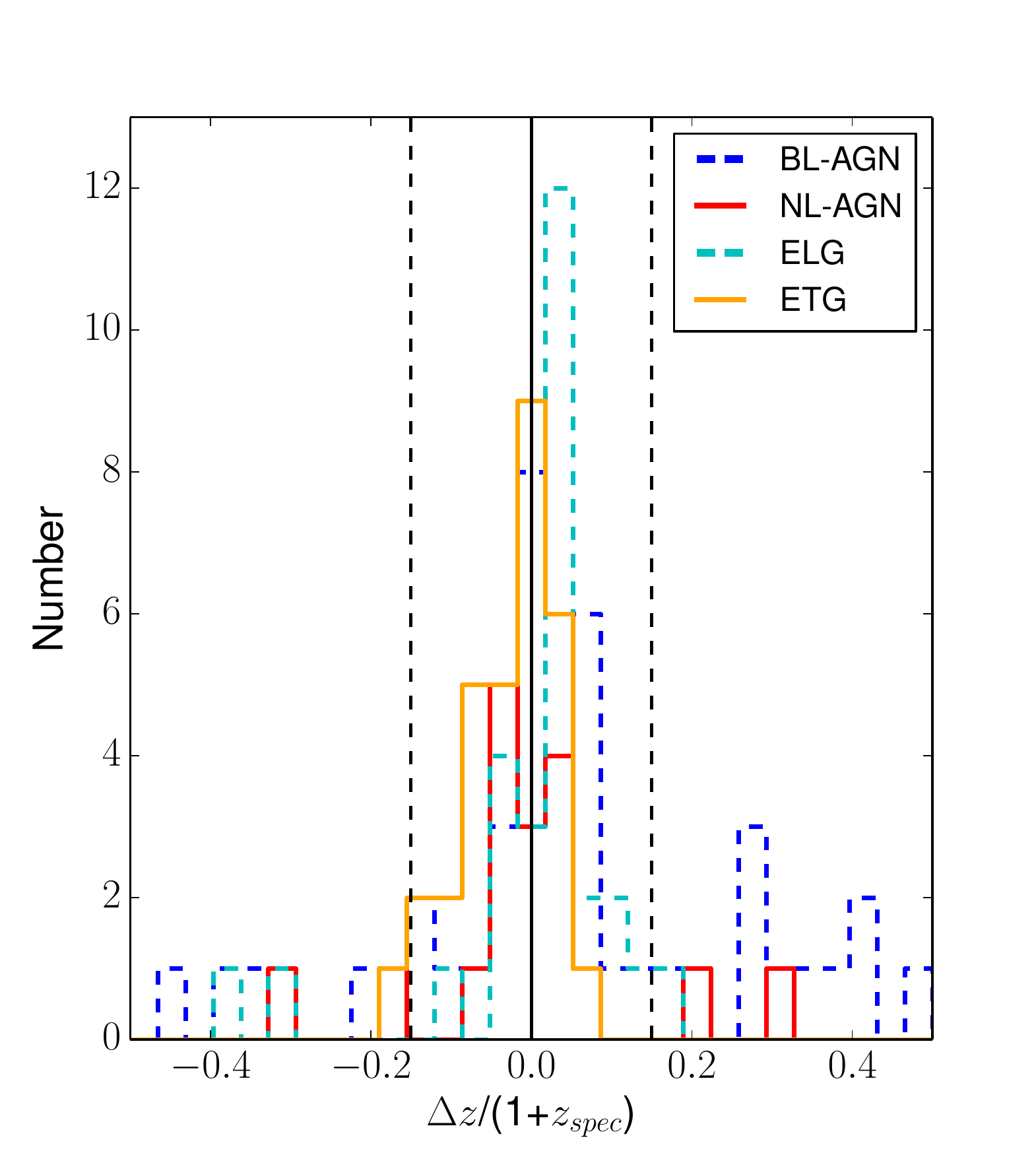} 
 \end{minipage} 
\begin{minipage}[b]{.45\textwidth} 
 \centering 
 \includegraphics[width=1.0\textwidth]{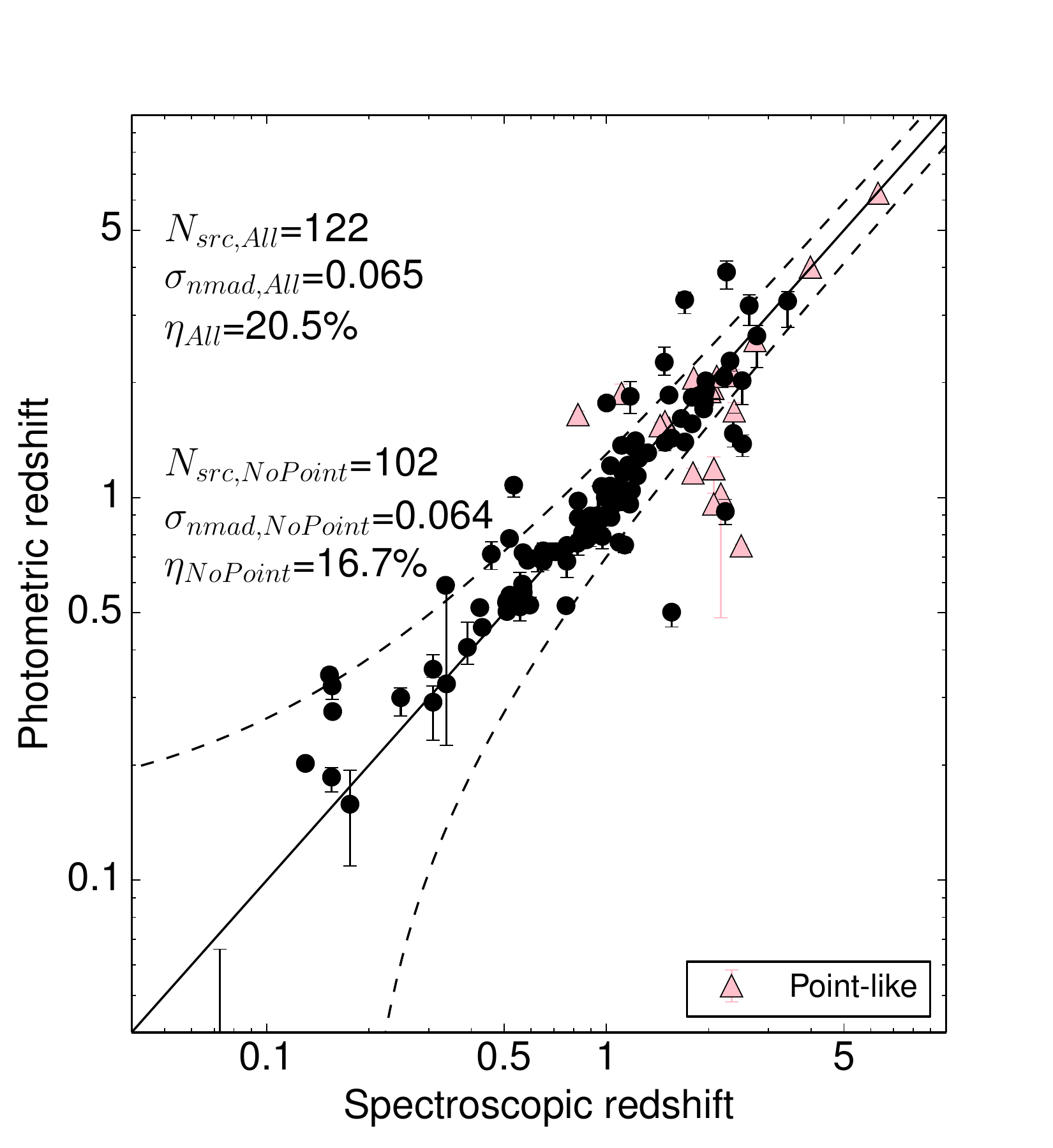} 
 \end{minipage} 
 \hspace{1.65cm}
  \begin{minipage}[b]{.45\textwidth} 
 \centering 
 \includegraphics[width=1.0\textwidth]{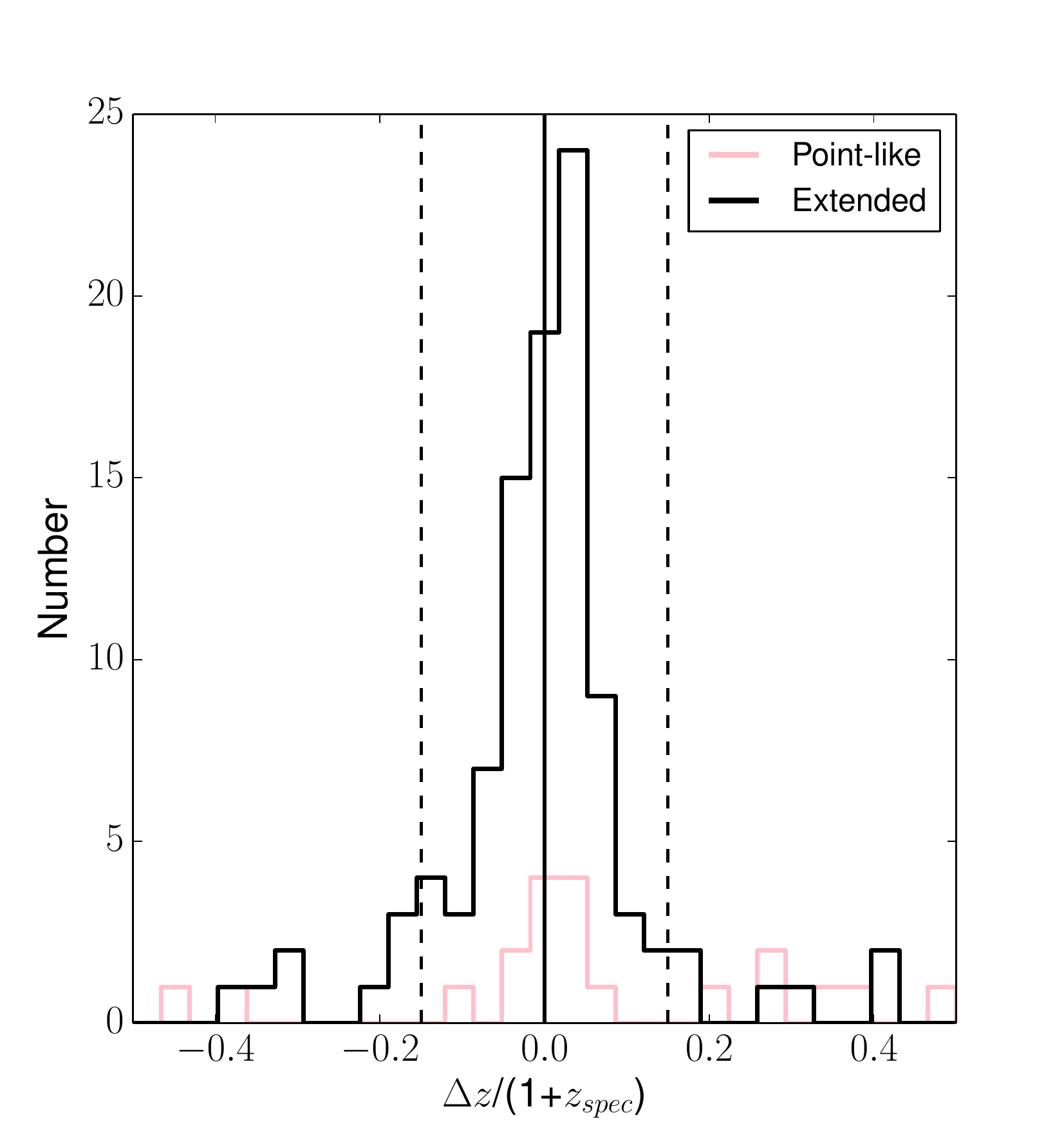} 
 \end{minipage} 
\caption{\normalsize 
\textit{Left}: Photometric redshift as a function of the spectroscopic one for the 122 J1030 sources with an available spectroscopic redshift. The black solid line marks the $z_{\rm spec}$=$z_{\rm phot}$ relation, while the dashed lines contain the $z_{\rm phot}$=$z_{\rm spec}\pm$0.15(1+$z_{\rm spec}$) region. Sources within this region are deemed to have a reliable photometric redshift. In the two panels, we highlight different classes of sources which are more likely to have a large fraction of outliers: spectroscopically classified BL-AGN (blue circles, top); morphologically-classified point-like sources (pink triangles, bottom).
\textit{Right}: Distribution of $\Delta z$/(1+$z_{\rm spec}$) for the sources reported in the left panel. The color code is the same of the two figures on the left.
}\label{fig:spec_vs_phot}
\end{figure*}

In Figure \ref{fig:spec_vs_phot} we show the distribution of the photometric redshifts as a function of the spectroscopic ones (left) and the distribution of $\Delta z$/(1+$z_{\rm spec}$) (right): in the top panels, we use different markers for each of the four spectroscopic classes we introduced in Section~\ref{sec:spec-z_class}, while in the bottom panels we highlight in pink the point-like sources. A visual comparison between the left panels clearly underlines how most point-like sources are BL-AGN. As it can be seen in the top panels, if we remove from our sample the BL-AGN population, we significantly improve both the accuracy $\sigma_{\rm NMAD,NoBL}$=0.060 and the outlier fraction, $\eta_{\rm NoBL}$=12.7\,\%. 

\begin{figure}[htbp]
 \centering
\includegraphics[width=1.\linewidth]{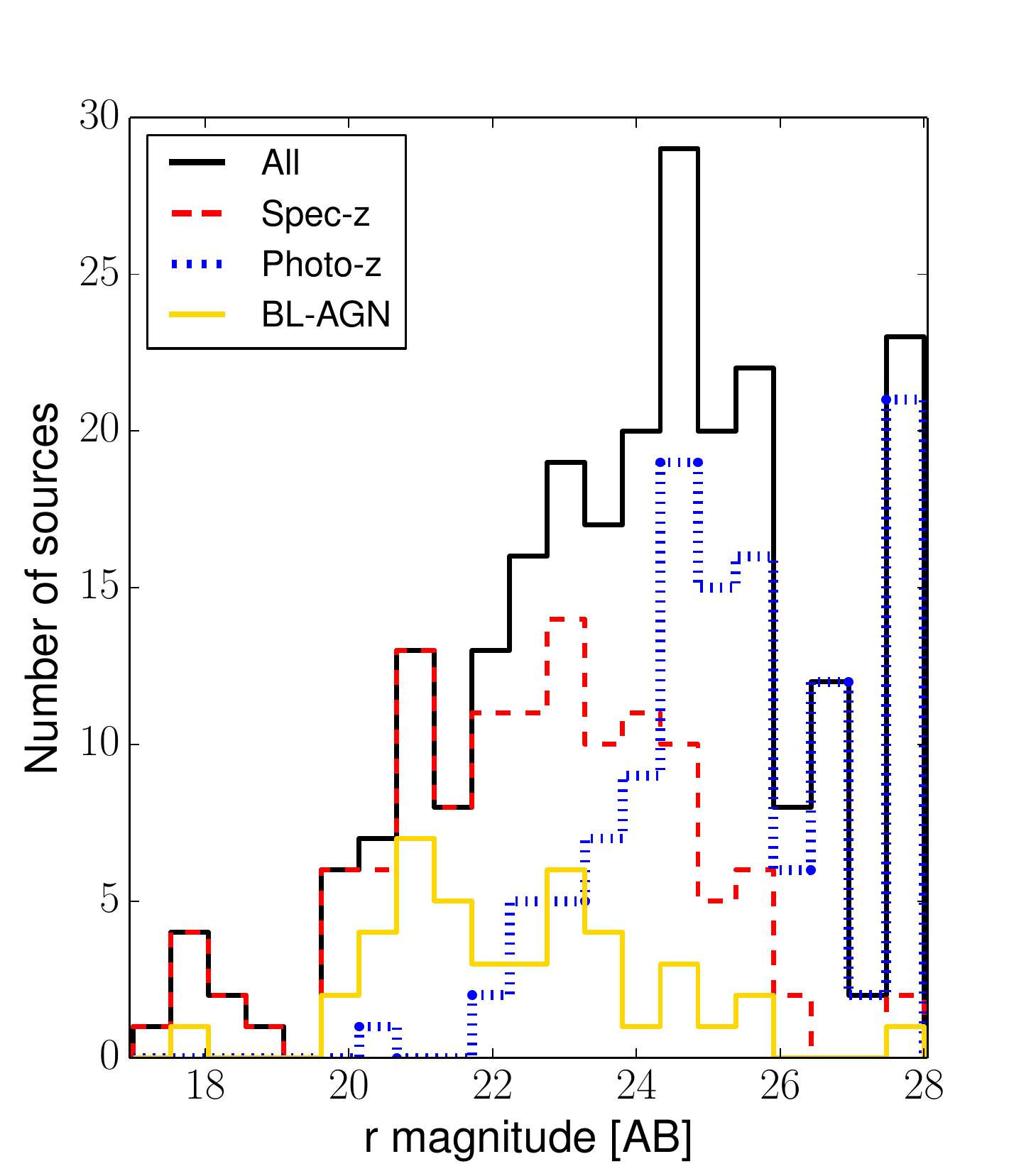}
\caption{\normalsize 
AB magnitude in the $r$ band of all the 243 extragalactic sources in the \cha\ J1030 sample having an optical/nIR counterpart (solid black line), the 123 sources with spectroscopic redshift (dashed red line), the 120 sources with only photometric redshift (dotted blue line) and the 43 broad-line AGNs (solid yellow line). We assign to the 24 sources with no r-band detection a magnitude r$_{\rm AB}$=27.5, which is their upper limit. The BL--AGN with an $r$--band upper limit at $r_{\rm aB}$=27.5 is the $z$=6.3 QSO SDSSJ1030+0525.
}\label{fig:histo_rmag}
\end{figure}

It is worth noting that the sample of 120 objects without a spectroscopic redshift includes only three point-like sources, and, more importantly, it is unlikely that it includes a significant number of BL-AGNs. As shown in Figure \ref{fig:histo_rmag}, BL-AGNs are, on average, significantly brighter than the sources with no spectroscopic redshift. More in detail, the average $r_{\rm AB}$ magnitude for BL-AGNs is $\langle r_{\rm AB,BL}\rangle$=22.1, and only one out of 43 BL-AGNs, the $z$=6.3 QSO SDSSJ1030+0525, is not detected with LBT/LBC in the $r$ band. Among the 120 sources without $z_{\rm spec}$, instead, 21 (18\,\%) do not have a $r$-band detection, and the average $r_{\rm AB}$ magnitude of the remaining 100 sources is $\langle r_{\rm AB,no-spec}\rangle$=24.9. We visually inspect the objects lacking $r$-band detection, and find that the lack of $r_{\rm AB}$ magnitude is not caused by artifacts or by issues in measuring the source flux. Therefore, these objects are highly likely to be either intrinsically faint, or heavily obscured, or at high redshift, or a combination of the three.

\subsection{Comparison with previous surveys}\label{sec:compare_other_works}
We report in Table \ref{tab:quality_comparison} the statistical properties of the photometric redshifts computed in several X-ray surveys. It can be seen that both $\sigma_{\rm NMAD}$ and the fraction of outliers, $\eta$, strongly depend on the number of photometric data-points used to perform the SED fitting. Consequently, surveys such as COSMOS, AEGIS-X or CANDELS/GOODS-S strongly benefit from years of multiwavelength imaging campaigns, and have smaller $\sigma_{\rm NMAD}$ and $\eta$ than our sample. Notably, all these surveys also have narrow- or intermediate-band magnitude information, which is key to properly measure photometric redshifts of unobscured, broad-line AGNs \citep[e.g.][]{ananna17}. 

The $\sigma_{\rm NMAD}$ and outlier fraction we obtain for J1030 are instead in good agreement with those obtained for the Stripe 82X wide-area survey \citep[$\sigma_{\rm NMAD}$=0.061, $\eta$=13.7\,\%][]{lamassa13a,ananna17} and those of the Lockman Hole survey \citep[$\sigma_{\rm NMAD}$=0.069, $\eta$=18.3\,\%][]{fotopoulou12}. Both surveys use a number of filters similar to ours (14 filters in the Stripe 82X survey and 21 in the Lockman Hole one), and lack narrow-- or intermediate--band information.

\begingroup
\renewcommand*{\arraystretch}{1.5}
\begin{table*}
\centering
\scalebox{1.}{
\vspace{.1cm}
 \begin{tabular}{cccccc}
 \hline
 \hline
 Survey      & Reference & Area   & N$_{\rm Band}$ (Narrow) & Accuracy & Outliers  \\
          &      & deg$^2$  &          &  $\sigma_{\rm NMAD}$  & $\eta$ \\
  \hline    
  J1030      & This work       & 0.09 & 21 (0)  & 0.065 & 20.5\,\%  \\
  J1030      & This work, no BL-AGN & 0.09 & 21 (0)  & 0.060 & 12.7\,\%  \\
  Lockman Hole  & \citet{fotopoulou12} & 0.13 & 21 (0)  & 0.069  & 18.3\,\% \\
  Stripe 82X   & \citet{ananna17}   & 31.3 & 14 (0)  & 0.061  & 13.7\,\% \\
  AEGIS-XD    & \citet{nandra15}   & 0.29 & 35 (18)   & 0.040 & 5.1\,\% \\
  CDF-S      & \citet{hsu14}     & 0.13 & 36 (18)  & 0.014 & 6.7\,\%  \\
  ECDF-S      & \citet{hsu14}     & 0.3  & 36 (18)  & 0.016 & 10.1\,\% \\
  \cha\ \leg   & \citet{marchesi16a}  & 2.2  & 38 (18) & 0.015  & 6.0\,\%  \\
  \hline
	\hline
\end{tabular}}
	\caption{\normalsize Normalized median absolute deviation ($\sigma_{\rm NMAD}$=1.48$\times$median(||$z_{\rm phot}$-$z_{\rm spec}$||/(1+$z_{\rm spec}$))) and fraction of photometric outliers $\eta$ (i.e., fraction of sources with ||$z_{\rm phot}$-$z_{\rm spec}$||/(1+$z_{\rm spec}$)$>$0.15 with respect to the overall number of sources with spectroscopic redshift) in different X-ray surveys. We also report the number of filters used to compute the photometric redshifts and, in parentheses, the number of narrow-band filters used in the computation.
	}
\label{tab:quality_comparison}
\end{table*}
\endgroup

\subsection{Comparison with Peca et al. (2021)}\label{sec:peca}
A subsample of sources analyzed in our work has been studied in a recent paper by our group \citep{peca21}. That work was based on the analysis of 54 obscured AGN candidates selected through their X-ray hardness ratio (HR$>$-0.1, where HR=$\frac{H-S}{H+S}$, and H and S are the net count rates in the 2--7\,keV and 0.5--2\,keV, respectively). These objects also had more than 30 net counts detected in the 0.5--7\,keV band, thus making it possible to perform an X-ray spectral fit. \citet{peca21} estimated, for the majority of the sources in the sample, both a photometric redshift (in 46 objects) and an X-ray redshift (38 objects).

In \citet{peca21}, the photometric redshifts were computed using \texttt{Hyperz}, as we do in this work. However, neither the IRAC CH3 and CH4, nor the HST data points were included in the SEDs. Furthermore, the objects in the \citet{peca21} sample were all obscured AGN candidates, where the optical emission is usually dominated by the host galaxy. For this reason, only the 75 ``galaxy'' templates were used to perform the SED fitting, without including in the analysis any ``AGN'' template.

Out of 46 \citet{peca21} sources with a photometric redshift, 31 (67\,\%) are in agreement with those we compute in our work. The agreement is assessed by following the same criterion we used to measure the agreement between spectroscopic and photometric redshifts in Section \ref{sec:spec_vs_phot} (i.e.,  ||$z_{\rm phot,M21}$-$z_{\rm phot,P21}$||/(1+$z_{\rm phot,M21}$)$<$0.15, where $z_{\rm phot,M21}$ is the photometric redshift computed in this work and $z_{\rm phot,P21}$ is the one derived in \citealt{peca21}). The remaining 15 sources are ``outliers''; notably, 10 of them are best-fitted with an AGN SED template, which is also the most likely cause of the discrepancy between the two photometric redshifts. It is also worth noting that five out of the 15 outliers also have a spectroscopic redshift: in four out of five cases, the agreement between photometric and spectroscopic redshift improves using the photometric redshifts presented in this paper.

To get an overall understanding of the accuracy of the photometric redshifts in the two works, we selected the 23 out of 46 sources that also have a spectroscopic redshift. For this subsample, our photometric redshifts have $\sigma_{\rm NMAD,M21}$=1.48$\times$median(||$z_{\rm phot}$-$z_{\rm spec}$||/(1+$z_{\rm spec}$))=0.062, while for the same objects \citet{peca21} have $\sigma_{\rm NMAD,P21,phot}$=0.090. The better accuracy of the photometric redshifts reported in this work can be explained with both the additional AGN templates used in our analysis and the fewer number of photometric data points used in \citet{peca21}.

Finally, we measured the agreement between our spectroscopic redshifts and the X-ray redshifts computed in \citet{peca21}: out of 19 objects with both $z_{\rm spec}$ and $z_{\rm X}$, 13 have $\Delta z$/(1+$z_{\rm spec}$)$<$0.15. The sample has $\sigma_{\rm NMAD,P21,X}$=0.150. We also note that for three out of six outliers the X-ray redshift is in agreement with the spectroscopic one within its 90\,\% confidence uncertainty: as already discussed in \citet{peca21}, these results confirm that in obscured AGN lacking a spectroscopic redshits, X-ray redshifts can provide a reliable estimate of the objects distance.

\subsection{Comparison with \texttt{EAzY}}
To perform a consistency check on the quality of our photometric redshifts, we recomputed them using the \texttt{EAzY} code \citep{brammer08}. We followed the same procedure we used with \texttt{Hyperz} and presented in Section~\ref{sec:SED_fitting}: more in detail, we fitted each SED twice, first with galaxy-only templates, then with hybrids which include AGN contribution. We then chose between the two best-fit solutions using the same flowchart reported in Figure~\ref{fig:flow_template}. We also used the K-band prior described in \citet{brammer08}.

Out of the 243 \cha\ J1030 extragalactic sources, 190 have photometric redshifts in agreement between the two codes, where the agreement is assessed using the criterion $\Delta z$/(1+$z_{\rm Hyperz}$)$<$0.15. The fraction of agreeing photometric redshifts is therefore $\sim$78\,\%, with $\sigma_{\rm NMAD}$=0.067. These results suggest that the \texttt{Hyperz}--computed photometric redshifts are reliable: as a reference, in a recent work on the eROSITA Final Equatorial-Depth Survey \citep{salvato21} the photometric redshifts were computed using both LePHARE and the machine learning-based method DNNz, and the outlier fraction was 38.4\,\%.

\begin{figure*} 
\begin{minipage}[b]{.47\textwidth} 
 \centering 
 \includegraphics[width=1.0\textwidth]{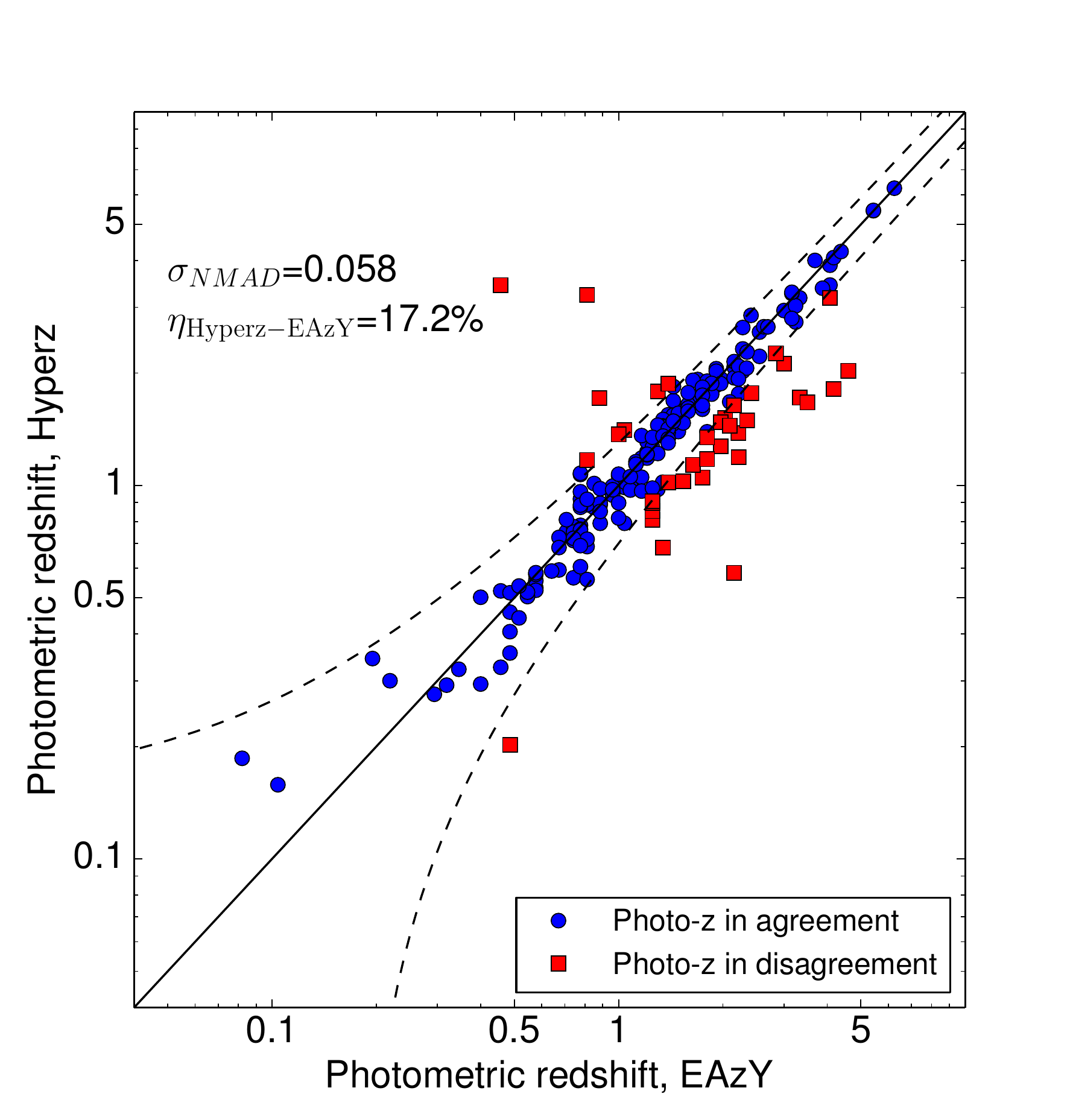} 
 \end{minipage} 
 \begin{minipage}[b]{.46\textwidth} 
 \centering 
 \includegraphics[width=1.0\textwidth]{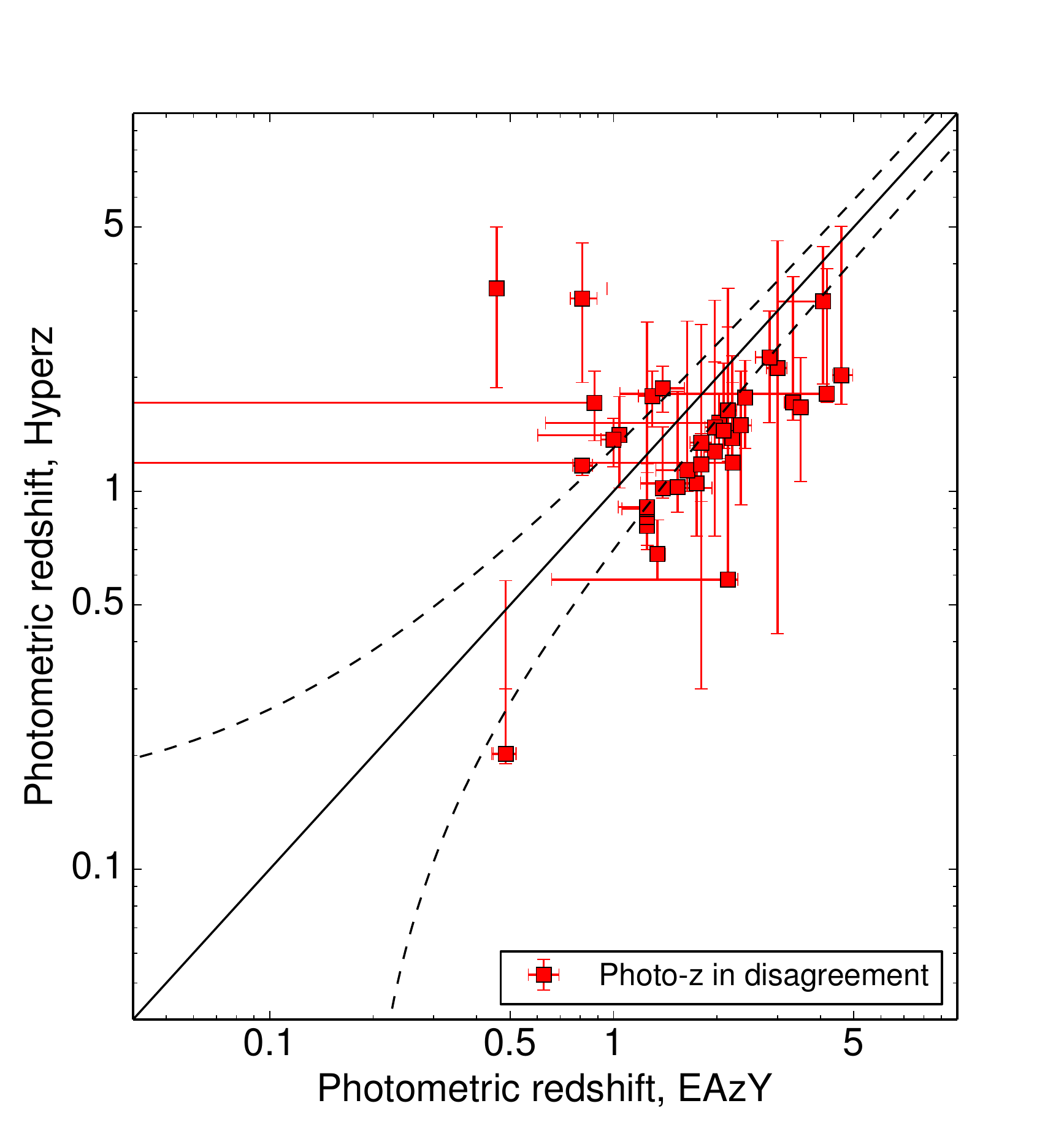} 
 \end{minipage} 
\caption{\normalsize 
\textit{Left}: Photometric redshifts computed with \texttt{Hyperz} as a function of those computed with \texttt{EAzY}. The black solid line marks the $z_{\rm EAzY}$=$z_{\rm Hyperz}$ relation, while the dashed lines contain the $z_{\rm EAzY}$=$z_{\rm Hyperz}\pm$0.15(1+$z_{\rm Hyperz}$) region. Sources where $\Delta z$/(1+$z_{\rm Hyperz})<$0.15 are plotted as blue circles, while sources where $\Delta z$/(1+$z_{\rm Hyperz})>$0.15 are plotted as red squares. \textit{Right}: Same as left, but with only the sources where $\Delta z$/(1+$z_{\rm Hyperz}>$0.15) and their 68\,\% uncertainties.
}\label{fig:hyperz_vs_EAzY}
\end{figure*}

Furthermore, in our work the subsample with photometric redshifts in disagreement is significantly populated by sources that are detected only in the K, IRAC or HST bands, and therefore have very large uncertainties and generally flat PDZ. When excluding these 21 sources from the computation, the fraction of objects with photometric redshifts in disagreement between \texttt{Hyperz} and \texttt{EAzY} drops to 17\,\% (38 outliers out of 222 sources), with $\sigma_{\rm NMAD}$=0.058. In Figure~\ref{fig:hyperz_vs_EAzY} we report the photometric redshifts computed with \texttt{Hyperz} as a function of those computed with \texttt{EAzY}, excluding the 21 sources with low data quality that we mentioned above. As it can be seen, most of the outliers are just outside the $\Delta z$/(1+$z$)=0.15 threshold (left panel), and many sources have photometric redshifts in agreement within the 68\,\% uncertainties (right panel). We note that here no $a$ correction is applied to the \texttt{EAzY} PDZs, thus the x-axis errors should be treated as conservative lower limits.

Finally, we also tested the ``multi-template'' option available within \texttt{EAzY}, which allows one to fit a combination of two or more templates among those available in the template library. We did not find any significant improvement in our photometric redshifts (i.e., we did not find an improvement neither in the fraction of photometric redshifts in agreement with the spectroscopic ones, nor in the fraction of \texttt{EAzY} photometric redshifts in agreement with the \texttt{Hyperz} ones). This result is not unexpected, given that many of the templates we use in this work are already a combination of a galaxy and an AGN one, therefore it was likely that the need for further combinations of templates would have been marginal.

In summmary, the results we obtained with \texttt{EAzY} strenghten the reliability of the \texttt{Hyperz} photometric redshifts. In the rest of the paper, we will therefore make use of the \texttt{Hyperz}--computed photo-$z$.

\section{The \cha\ J1030 multiwavelength catalog}\label{sec:catalog_description}
The \cha\ J1030 multiwavelength catalog is available online\footnote{\url{http://j1030-field.oas.inaf.it/chandra_1030}}, in fits format. Here we describe the catalog columns.

\begin{itemize}
    \item \textit{Column 1.} X-ray source ID, from \citet{nanni20}.
    \item \textit{Columns 2–3}. X-ray coordinates of the source, from \citet{nanni20}.
    \item \textit{Column 4.} ID from the $z$-detected LBC catalog \citep{morselli14}. Sources detected in the $r$-deep catalog only are flagged with a $r$, sources detected only in the IRAC band are identified as ``IRAC NNN'', where NNN is a numerical ID.
    \item \textit{Columns 5–6}.  Coordinates of the source identified as the counterpart of the X-ray source in \citet{nanni20}.
    \item \textit{Columns 7--10}: Counterpart magnitude AUTO in the $r$, $z$, $J$, and 4.5\,$\mu$m bands, respectively, as reported in \citet{nanni20}.
    \item \textit{Column 11.} Redshift of the source. If a spectroscopic redshift is available, we select it as the source redshift; otherwise, the photometric redshift is reported.
    \item \textit{Column 12.} Spectroscopic redshift of the source, $z_{\rm spec}$. Sources with a spectrum lacking clearly identifiable features are flagged with -9.99; sources with no detection in a slit are flagged with -8.88; and sources for which no spectroscopic observations were performed are flagged with -99.
    \item \textit{Column 13.} Spectral quality flag. 0--Sources without $z_{\rm spec}$; 1--Sources with uncertain redshift; 2--Sources with secure redshift.
    \item \textit{Column 14.} Spectral type. 
    \item \textit{Column 15.} Spectral origin. D--DEIMOS on Keck II; F--FORS2 on VLT; L--LUCI on LBT; M--MODS on LBT; Mu--MUSE on VLT; S--SDSS.
    \item \textit{Column 16.} Photometric redshift.
    \item \textit{Columns 17--18.} 68\,\% confidence level lower and upper boundaries of the photometric redshift, computed after applying the $\sigma_{\rm corr}$ correction presented in Section \ref{sec:SED_fitting}.
    \item \textit{Column 19.} Reduced $\chi^2$ of the photometric redshift.
    \item \textit{Column 20.} Template class of the photometric redshift (AGN or galaxy).
    \item \textit{Column 21.} Source morphology (point-like or extended), determined using the LBC \hbox{$z$-band} image.
    \item \textit{Column 22.} Number of photometric points in the spectral energy distribution of the source. Upper limits are not included.
    \item \textit{Columns 23--24.} Observed 0.5--2\,keV and 2--7\,keV fluxes, from \citet{nanni20}.
    \item \textit{Columns 25--27.} Hardness ratio (HR=$\frac{H-S}{H+S}$, where H and S are the net count rates in the 2--7\,keV and 0.5--2\,keV, respectively) and corresponding 1\,$\sigma$ lower and upper boundaries, as reported in \citet{nanni20}. In sources where the lower and upper boundaries are equal to the best-fit value, the HR is reported as the 3\,$\sigma$ upper (lower) limit.
    \item \textit{Column 28.} Neutral Hydrogen column density (N$_{\rm H}$) derived using the source hardness ratio and redshift \citep[see, e.g., Appendix B in][for a detailed description of this approach]{peca21}, assuming a photon index $\Gamma$=1.8, as typically seen in AGN X-ray spectra \citep{marchesi16c}.
    \item \textit{Columns 29--30.} Neutral Hydrogen column density (N$_{\rm H}$) 1\,$\sigma$ lower and upper boundaries, derived using the source redshift, and the HR 1\,$\sigma$ lower and upper boundaries, assuming a power-law photon index $\Gamma$=1.8. In sources where the lower and upper boundaries are equal to the best-fit value, the column density is a 3\,$\sigma$ upper limit.
    \item \textit{Columns 31--32.} Rest-frame, absorption-corrected 0.5--2\,keV and 2--7\,keV luminosities, obtained using the 0.5--2\,keV and 2--7\,keV observed fluxes and the column density N$_{\rm H}$, and assuming a power-law photon index $\Gamma$=1.8.
    \item \textit{Columns 33--34.} Detection flag in the 0.5--2\,keV and 2--7\,keV. `1'' means that the source is detected in a given band; ``-1'' that the source is not detected. If a source is not detected in a given band, the reported flux and luminosity values in the same band are 3\,$\sigma$ upper limits.
\end{itemize}

\section{Main properties of the J1030 extragalactic population}\label{sec:sample_properties}
\subsection{Redshift and luminosity distribution}
\begin{figure}[htbp]
 \centering
\includegraphics[width=0.98\linewidth]{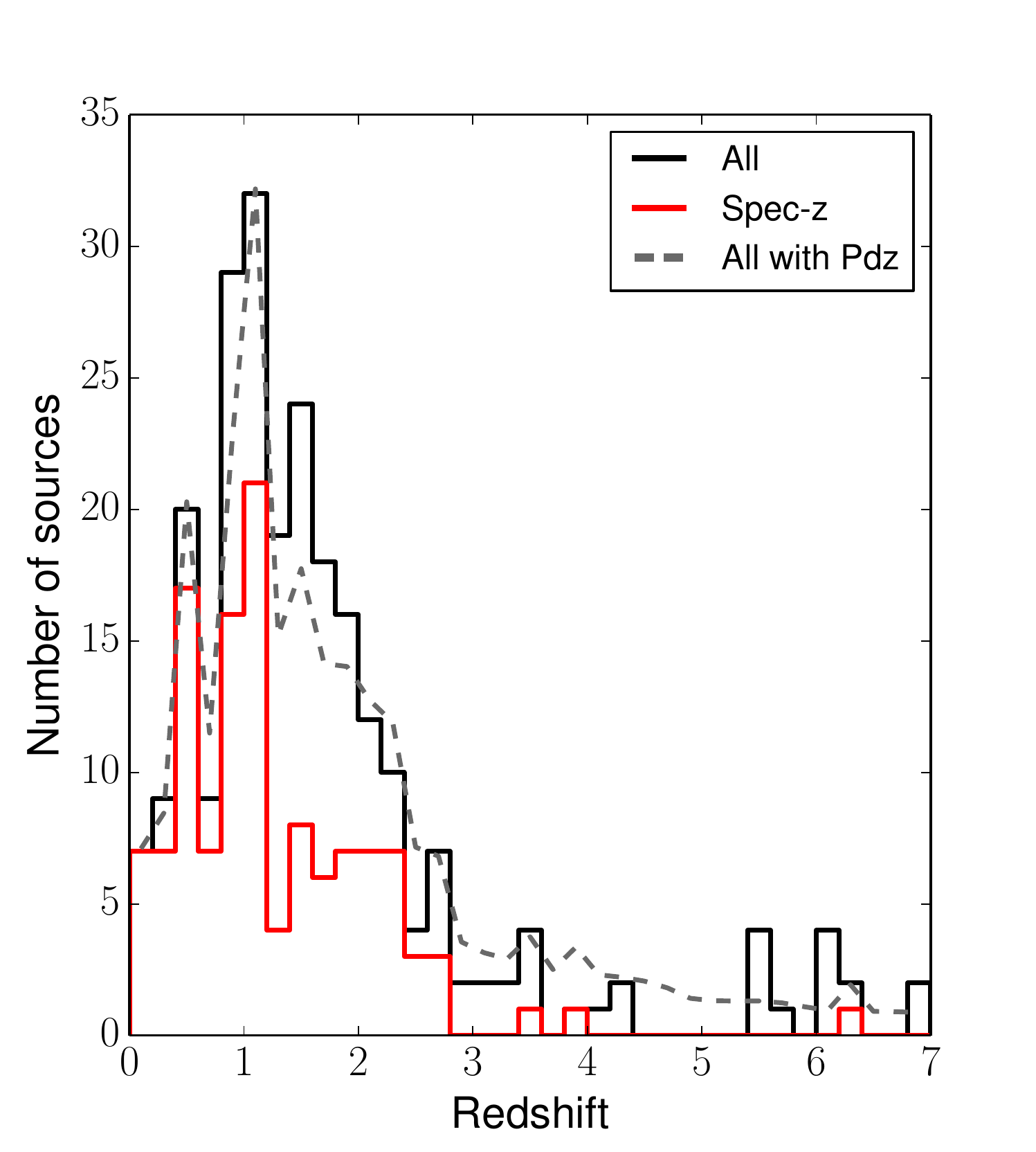}
\caption{\normalsize 
Redshift distribution of the 243 \cha\ J1030 sources with an optical and/or IR counterpart. The black solid line shows the overall distribution computed using the best-fit redshift value, while the grey dashed line shows the distribution obtained by summing all the redshift probability distribution function values (for photometric redshifts; spectroscopic redshifts are assumed to have PDZ=1 at their $z_{\rm spec}$ value). The spectroscopic redshift distribution is shown as a red solid line.
}\label{fig:histo_z_phot}
\end{figure}

In Figure \ref{fig:histo_z_phot} we report the redshift distribution of the \cha\ J1030 extragalactic sample. We plot as a black solid line the distribution of the ``best'' redshifts, which are the spectroscopic redshifts for the 123 sources with $z_{\rm spec}$, and the photometric redshifts for the remaining 120. We instead use a gray dashed line to plot the sum of the redshift probability distribution functions: if a source has a spectroscopic redshift, we assume PDZ=1 at the $z_{\rm spec}$ value. Finally, the spectroscopic redshifts distribution is shown as a red solid line.  As it can be seen, in the redshift range $z$=[0--3] there is a good agreement between the overall distribution computed using the spectroscopic redshifts (when available) and the best-fit $z_{\rm phot}$ values and the one obtained using $z_{\rm spec}$ (when available) and the PDZs. This is due both to the fact that at low redshift the spectroscopic completeness is larger and that photometric redshifts up to {\it z}$\,\sim\,$3 have, on average, narrower PDZs than high-$z$ sources.

\begin{figure*} 
\begin{minipage}[b]{.48\textwidth} 
 \centering 
 \includegraphics[width=1.\textwidth]{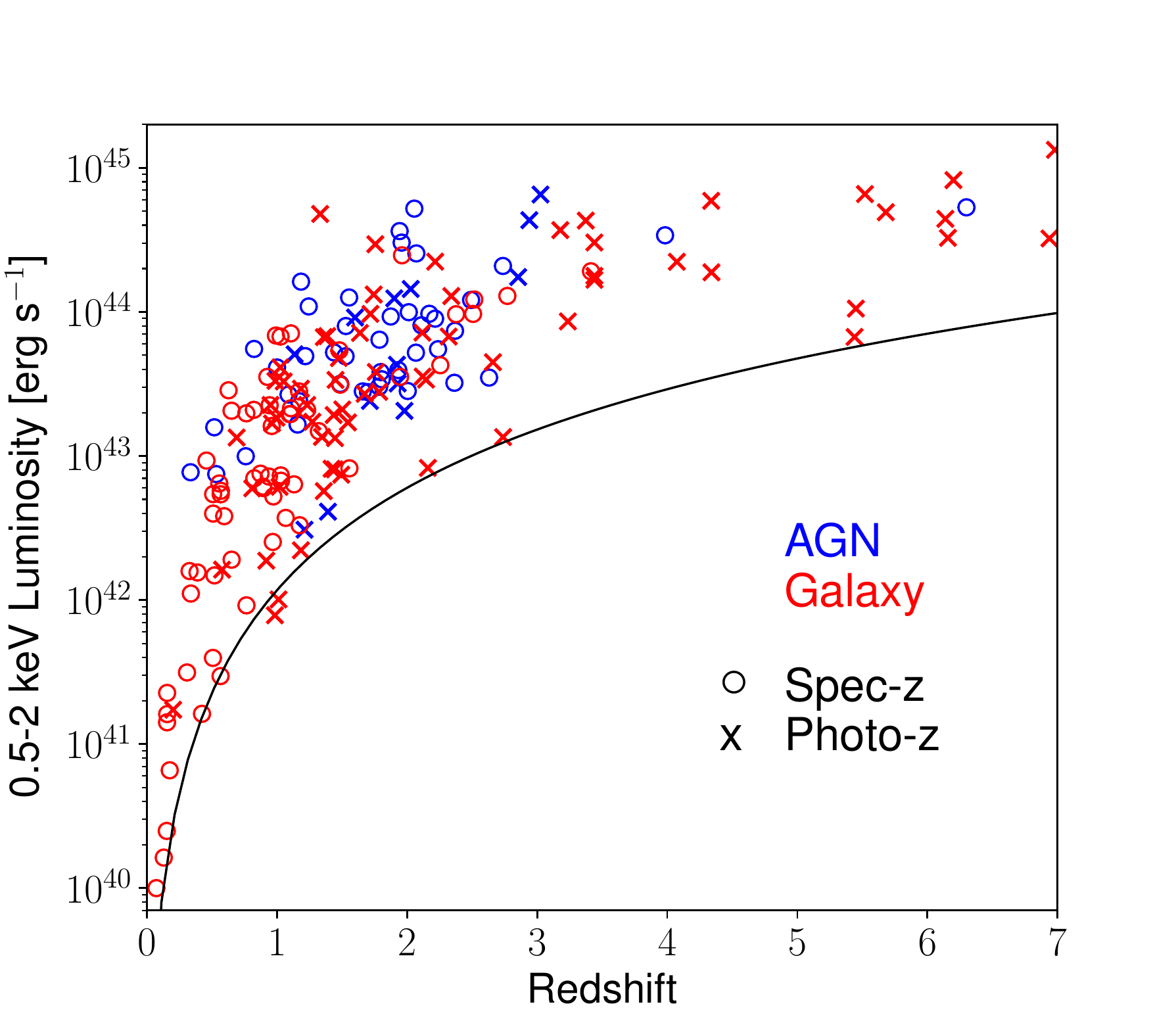} 
 \end{minipage} 
 \hspace{0.1cm}
\begin{minipage}[b]{.49\textwidth} 
 \centering 
 \includegraphics[width=1.\textwidth]{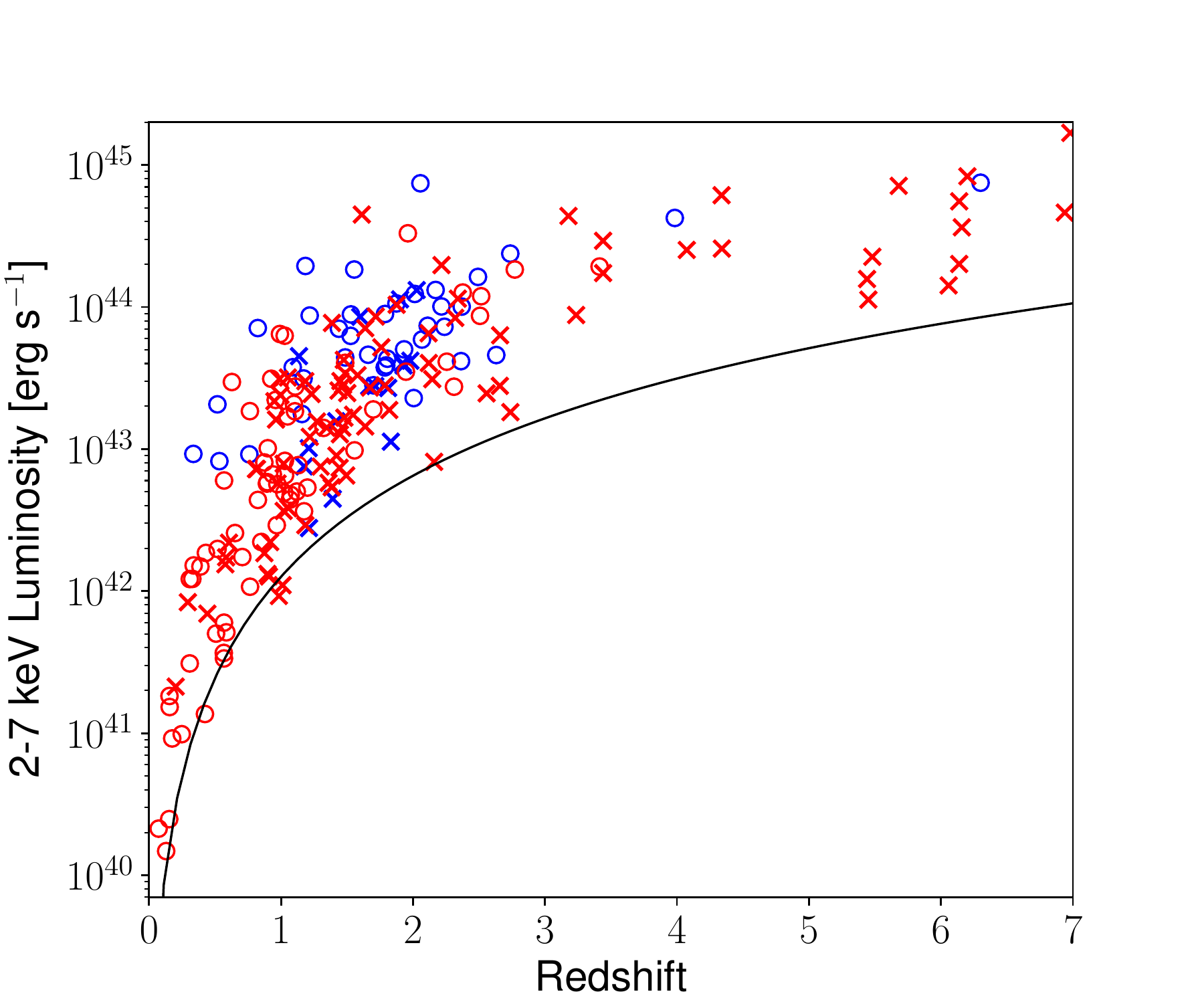} 
 \end{minipage} 
 \begin{minipage}[b]{.49\textwidth}
 \centering 
 \includegraphics[width=1.\textwidth]{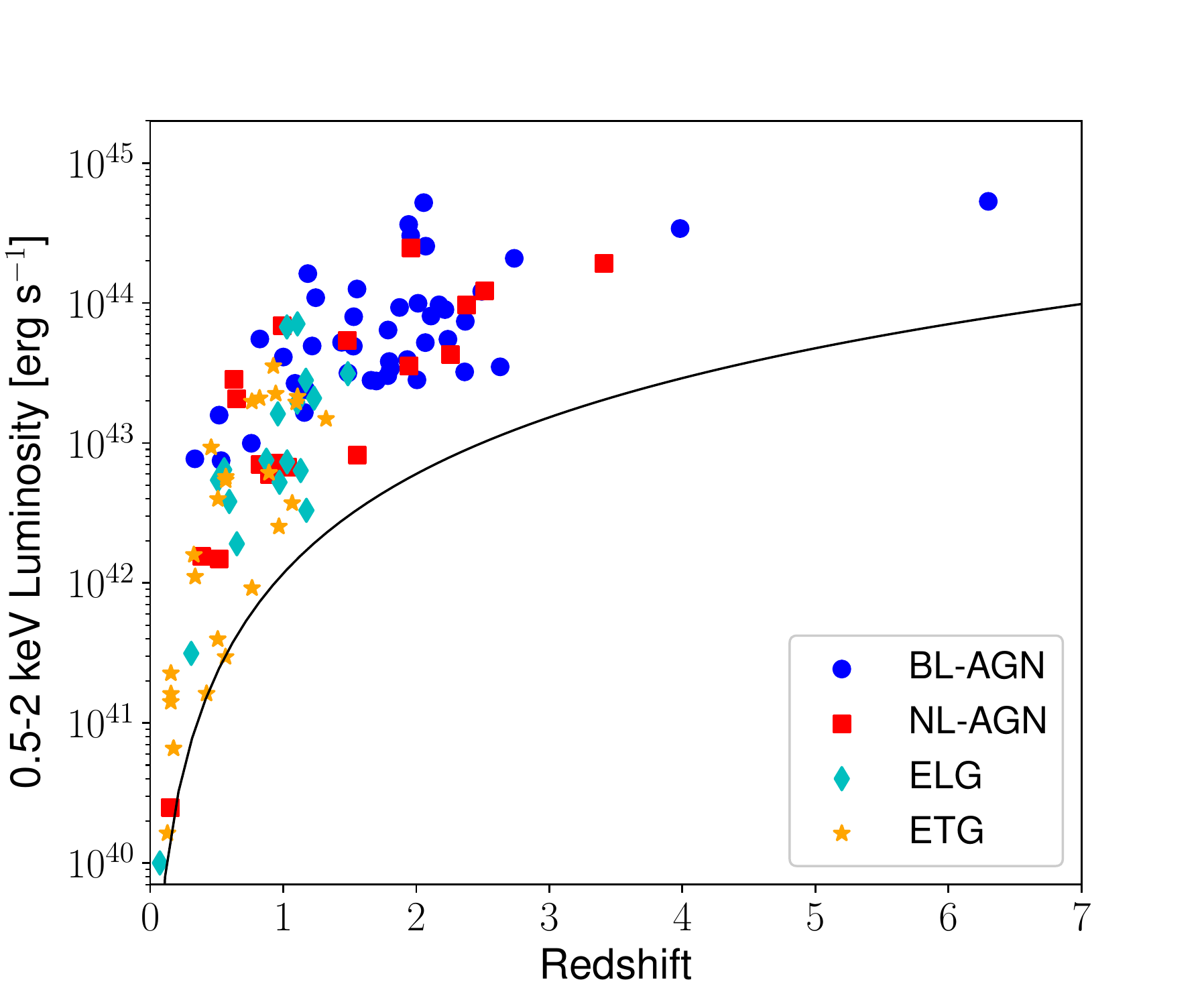} 
 \end{minipage} 
 \hspace{0.1cm}
\begin{minipage}[b]{.50\textwidth} 
 \centering 
 \includegraphics[width=1.\textwidth]{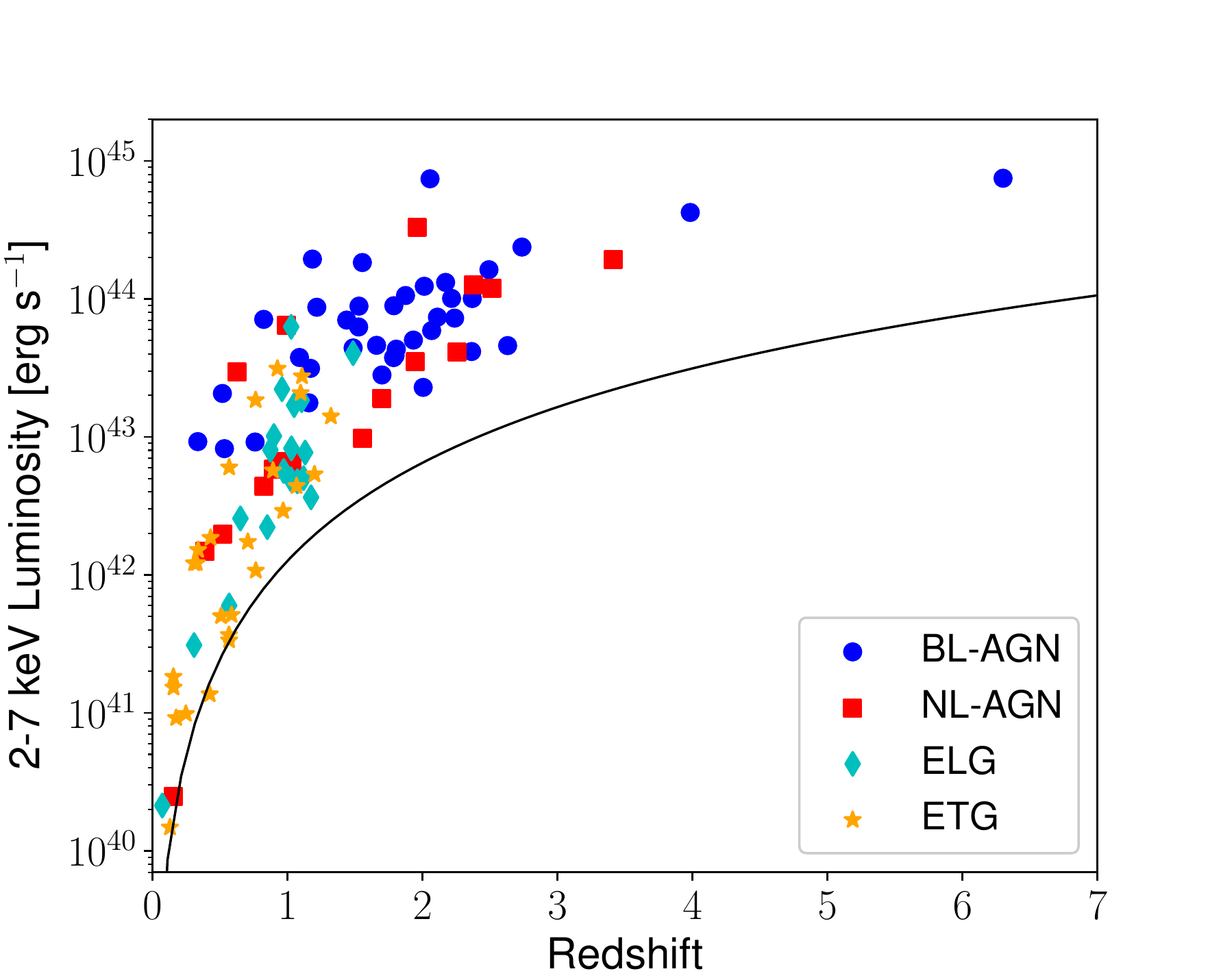} 
 \end{minipage}
\caption{\normalsize 
\textit{Top}: Intrinsic, absorption-corrected luminosity in the 0.5--2\,keV (left) and 2--7\,keV (right) band as a function of redshift. Sources with a spectroscopic redshift are plotted with circles, while sources with only a photometric redshift are plotted with crosses. Sources with a BL-- or NL-AGN spectrum or with a SED best-fitted with an AGN template are plotted in blue, while sources with a ETG or ELG spectrum, or with a SED best-fitted with a galaxy template are plotted in red. We also plot as a black solid line the luminosity corresponding to the flux at which 10\,\% of the field is covered.
\textit{Bottom}: same as above, but for the spectroscopic subsample alone. BL--AGNs are plotted as blue circles, NL--AGNs as red squares, emission line galaxies as cyan diamonds, and early type galaxies as orange stars.
}\label{fig:z_vs_lx}
\end{figure*}

In Figure \ref{fig:z_vs_lx} we report the intrinsic, absorption--corrected 0.5--2\,keV (left panel) and 2--7\,keV (right panel) luminosities as a function of redshift for the sources in our sample. We compute an estimate of the absorbing column density ($N_{\rm H}$) using the hardness ratio  (HR) value reported in \citet{nanni20} and the redshift information, following the approach discussed, for example, in \citet{marchesi16a}. In this computation, we assume a typical AGN power-law photon index $\Gamma$=1.8 \citep[see, e.g.,][]{marchesi16c}. We then use $N_{\rm H}$ to derive an absorption correction factor which we apply to the observed luminosity computed using the X-ray fluxes reported in \citet{nanni20}.

As expected, AGN--dominated sources (plotted in blue in Figure \ref{fig:z_vs_lx}, top panels) are more luminous and observed at higher redshifts than galaxy-dominated objects (plotted in red in the same Figure). More in detail, among the sources with a spectroscopic redshift BL-- and NL--AGN have median redshift $\mu_{\rm z-sp,AGN}$=1.79 and median intrinsic 2--7\,keV luminosity Log($\mu_{\rm L-2-7,sp,AGN}$)=43.8, while ELGs and ETGs have  $\mu_{\rm z-sp,Gal}$=0.93 and Log($\mu_{\rm L-2-7,sp,Gal}$)=42.8. This is further highlighted in the bottom panels of Figure~\ref{fig:z_vs_lx}, where we plot the 0.5--2\,keV and 2--7\,keV luminosity as a function of redshift for the subsample of sources with spectroscopic redshift: the different colors identify the different spectral classes we presented in Section~\ref{sec:spec-z_class}.

The difference between AGN-- and galaxy-- dominated sources is less significant for the subsample of sources that only have a photometric redshift: here, sources best-fitted with an AGN template have median redshift  $\mu_{\rm z-ph,AGN}$=1.62 and median intrinsic 2--7\,keV luminosity Log($\mu_{\rm L-2-7,ph,AGN}$)=43.4, while sources best-fitted with a galaxy template have $\mu_{\rm z-ph,gal}$=1.49 and median intrinsic 2--7\,keV luminosity Log($\mu_{\rm L-2-7,ph,gal}$)=43.4.

\subsection{X-ray sources as an efficient tracer of large-scale structures}\label{sec:structures}
\begin{figure*} 
\begin{minipage}[b]{.49\textwidth} 
 \centering 
 \includegraphics[width=1.0\textwidth]{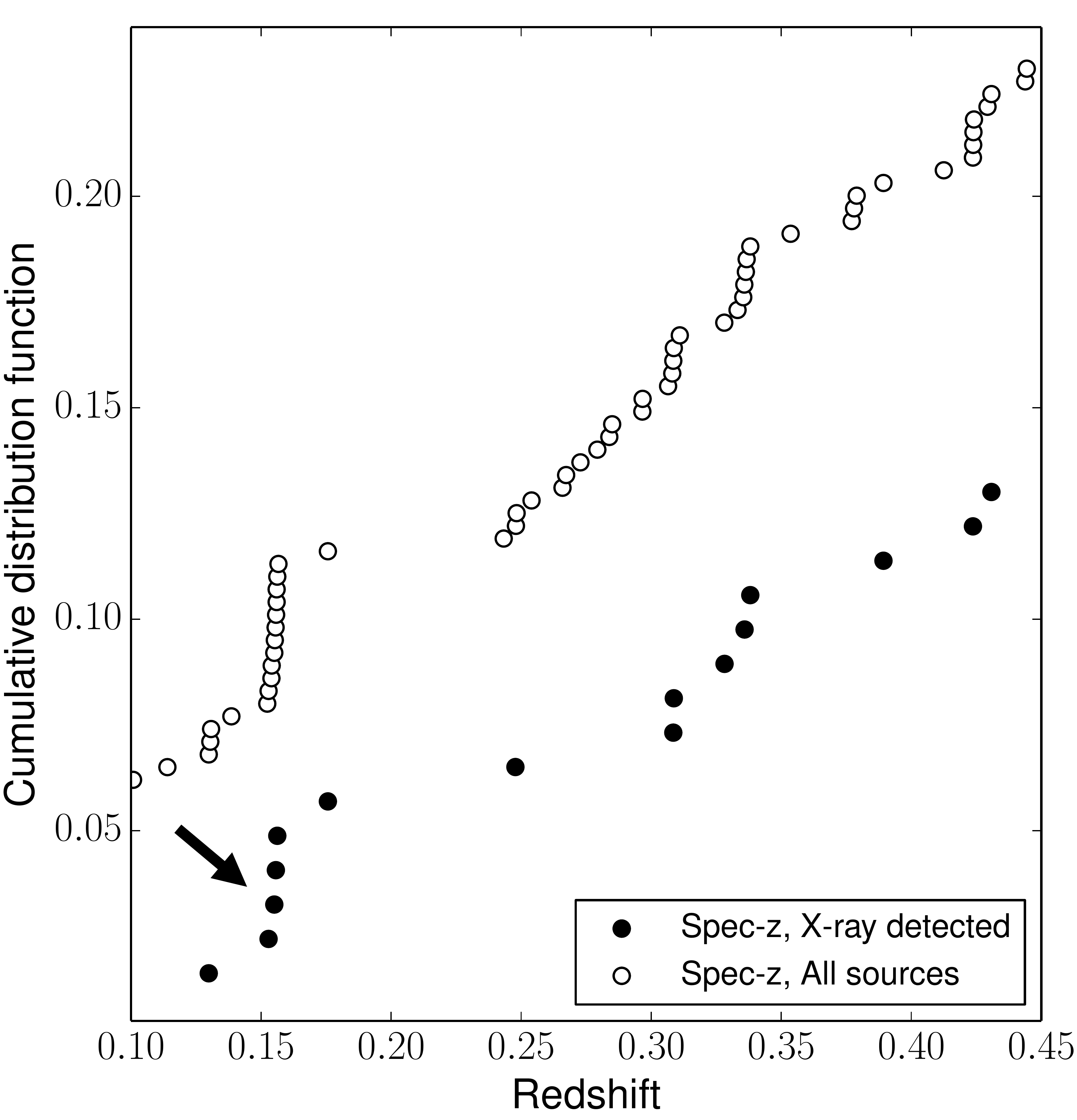} 
 \end{minipage} 
\begin{minipage}[b]{.49\textwidth} 
 \centering 
 \includegraphics[width=1.0\textwidth]{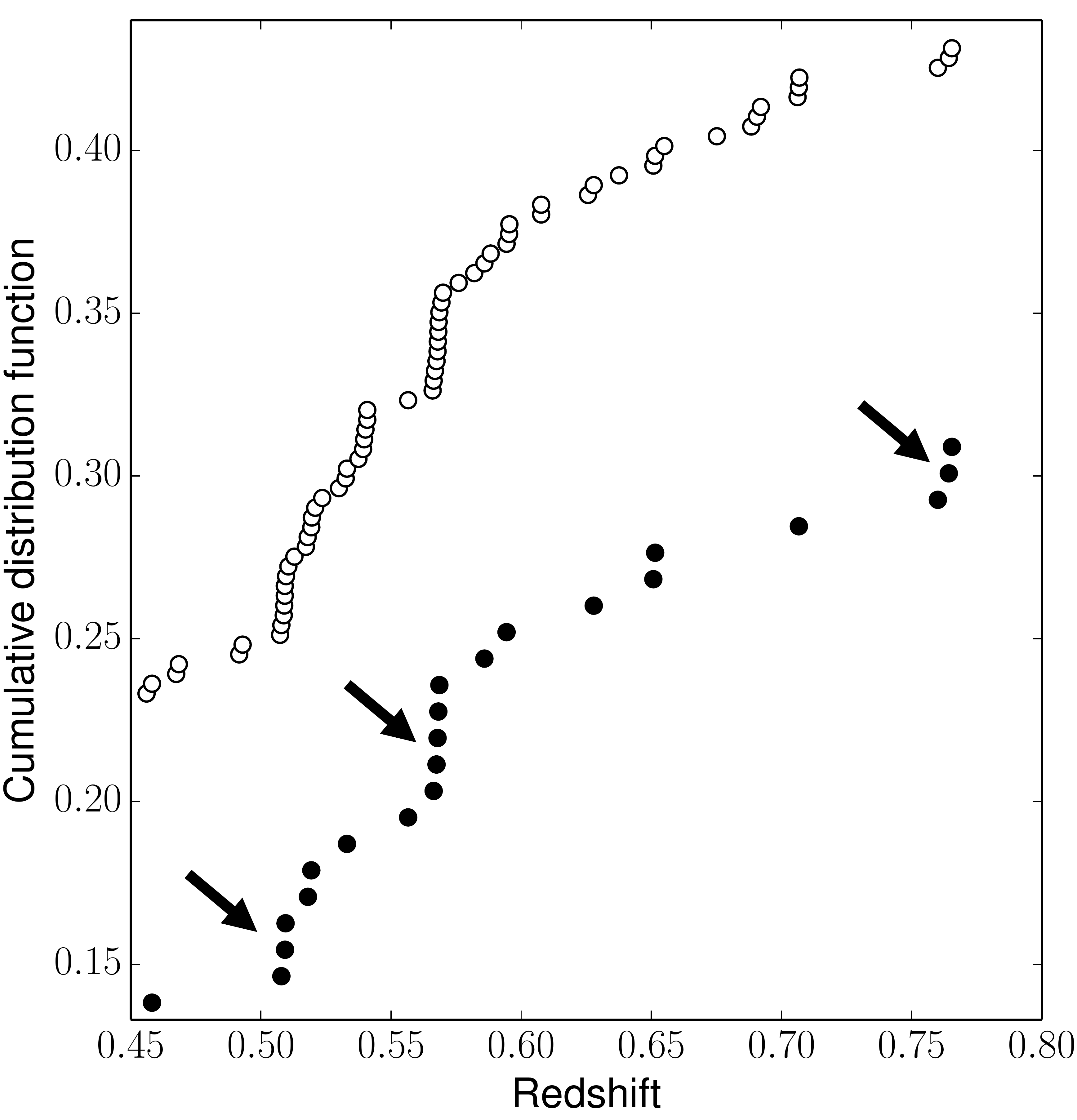} 
 \end{minipage} 
\begin{minipage}[b]{.49\textwidth} 
 \centering 
 \includegraphics[width=1.0\textwidth]{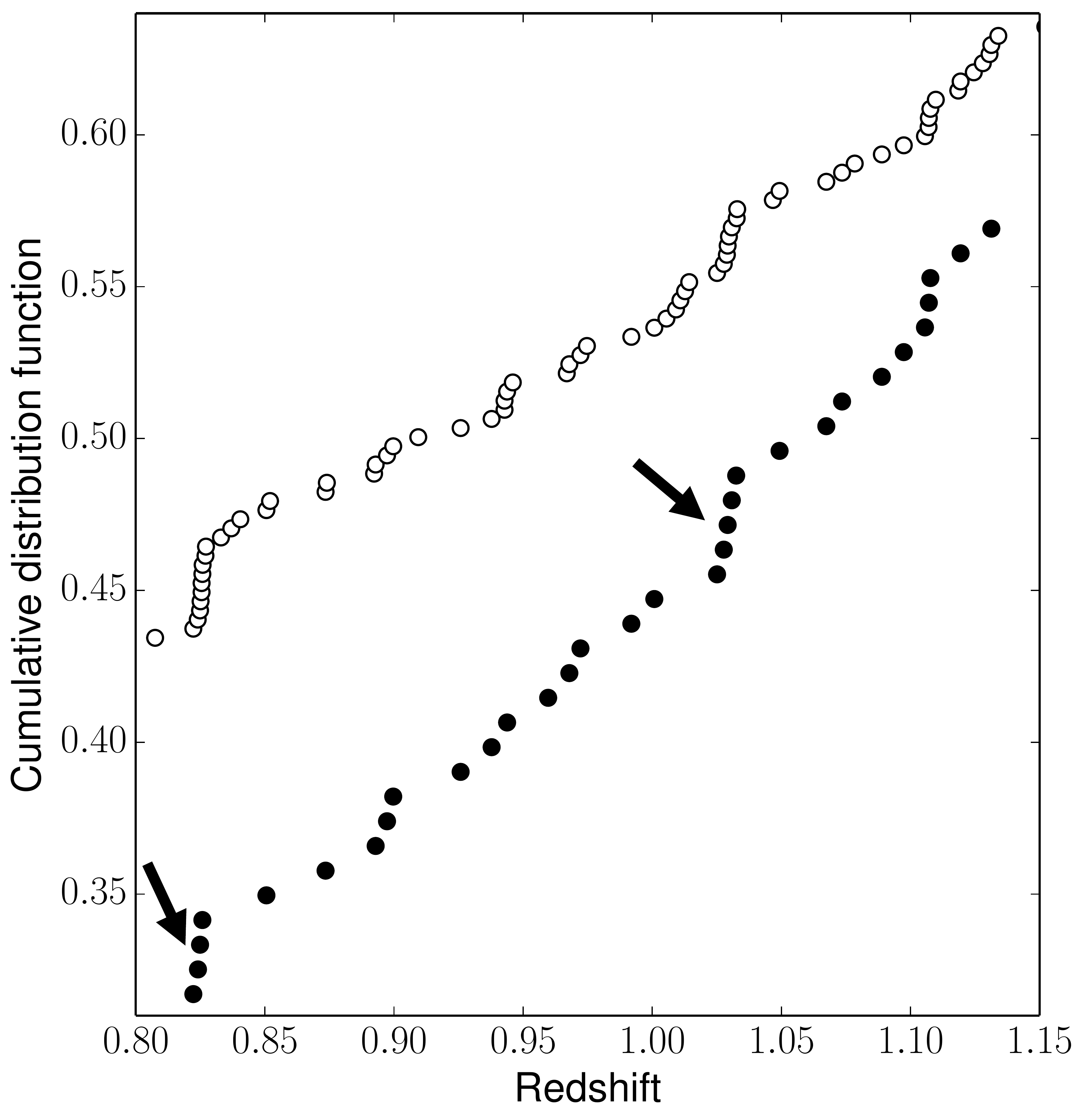} \end{minipage} \begin{minipage}[b]{.49\textwidth} 
 \centering 
 \includegraphics[width=1.0\textwidth]{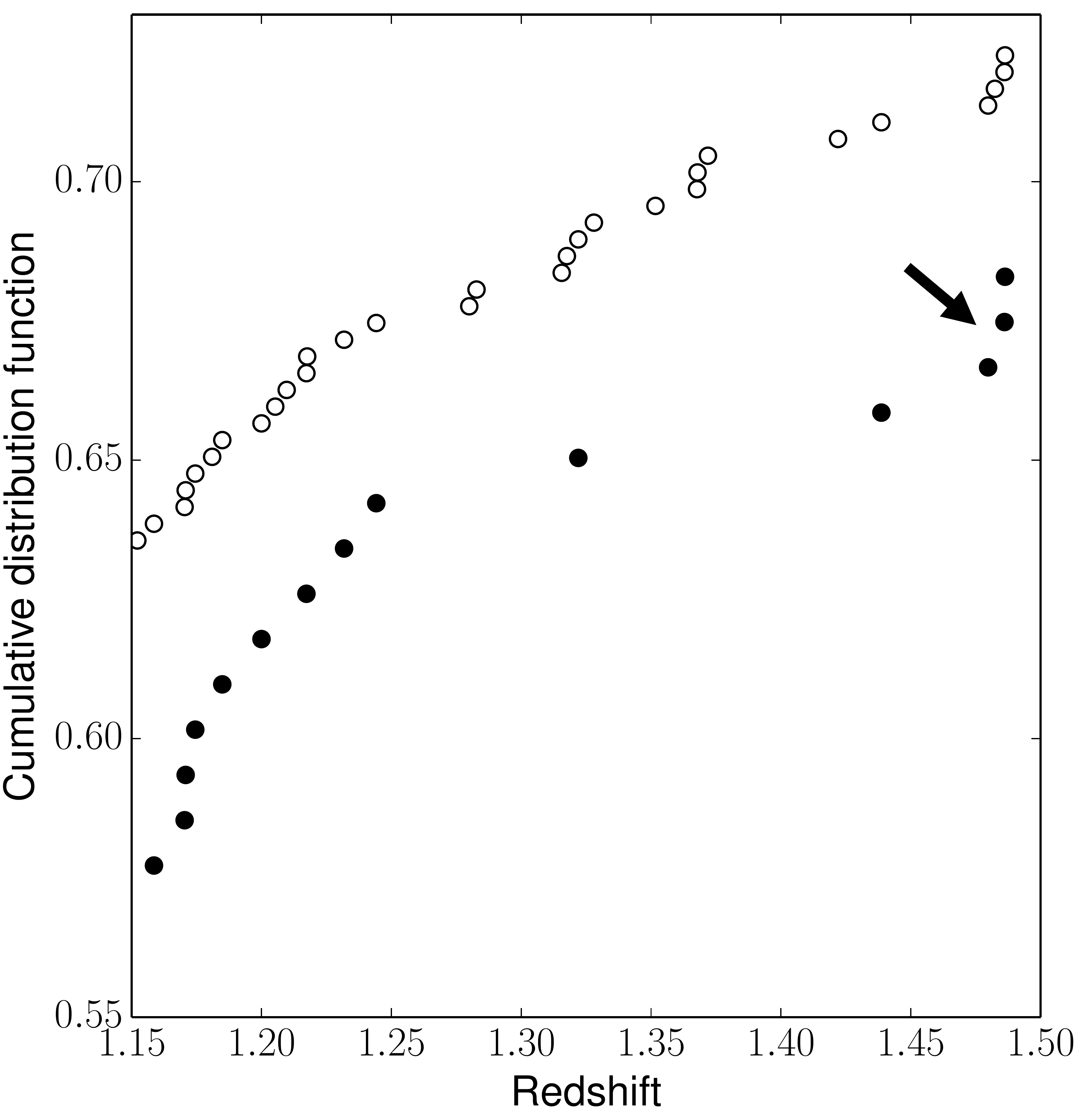} 
 \end{minipage} 
\caption{\normalsize 
Cumulative distribution function for the \cha\ J1030 sources with a spectroscopic redshift (full markers), and for all the J1030 sources with a spectroscopic redshift, including the non-X-ray detected ones (empty markers). The overall CDF distribution is shifted by 0.05 for visualization purposes. The structures identified within the \cha\ J1030 population are marked with arrows: as it can be seen, the X-ray sources effectively track the overall population.
}\label{fig:z_vs_cdf}
\end{figure*}

Following \citet{iwasawa20}, we plot in Figure \ref{fig:z_vs_cdf} the cumulative distribution function (CDF) of \cha\ J1030 spectroscopic redshifts up to $z$=1.5. Such a representation provides a good visualization of candidate redshift structures in the field, which readily appear as sharp CDF rises, and is free of binning issues. Similarly to what is commonly seen in deep spectroscopic surveys, several redshift structures are observed.

To assess the significance of these structures, we followed the same procedure adopted by \citet{gilli03} to identify redshift structures as traced by X-ray sources in the 1\,Ms Chandra Deep Field South. Sources were distributed in velocity space $V= c\,\rm{ln}(1+z)$ rather than in redshift space, since $dV$ corresponds to local velocity variations relative to the Hubble expansion. The signal velocity distribution was then smoothed with a Gaussian with $\sigma_S=300$\,km\,s$^{-1}$, to match the typical velocity dispersion observed in galaxy structures \citep[e.g.,][]{cohen99,gilli03}. The background velocity distribution was instead smoothed with a broader Gaussian, having $\sigma_B=1.5\times 10^4$\,km\,s$^{-1}$. We searched for candidate peaks in the signal distribution by computing their signal-to-noise ratio S/N = $(S-B)/\sqrt{B}$, where $S$ is the number of sources within $\Delta V= \pm1000$ km s$^{-1}$ around the center of each peak candidate, and $B$ is the number of background sources in the same interval. The value of $\Delta V$ was chosen to optimize the S/N values of poorly populated peaks. 

Adopting the thresholds $S\geq3$ (i.e., a peak must contain at least three sources to be statistically significant) and S/N $>$3.8, we found seven peaks in the signal velocity distribution. By means of detailed simulations, by analyzing the 1\,Ms CDF-S sample \citet{gilli03} demonstrated that the chosen detection thresholds ensure a fraction of spurious peak detections below 10\%. This is a reasonable estimate of the spurious peak fraction also for our sample, given that the number and distribution of the spectroscopic redshifts for the X-ray sources  in J1030 are both very similar to those of the 1\,Ms CDF-S X-ray sources. The corresponding redshifts of the peaks detected by our procedure are listed in Table~\ref{tab:spike_significance}, where we also list the number of objects $S$ detected in each peak and the Poisson probability of observing $S$ sources given the estimated background value $B$.

\begin{figure*} 
\begin{minipage}[b]{.63\textwidth} 
 \centering 
 \includegraphics[width=0.99\textwidth]{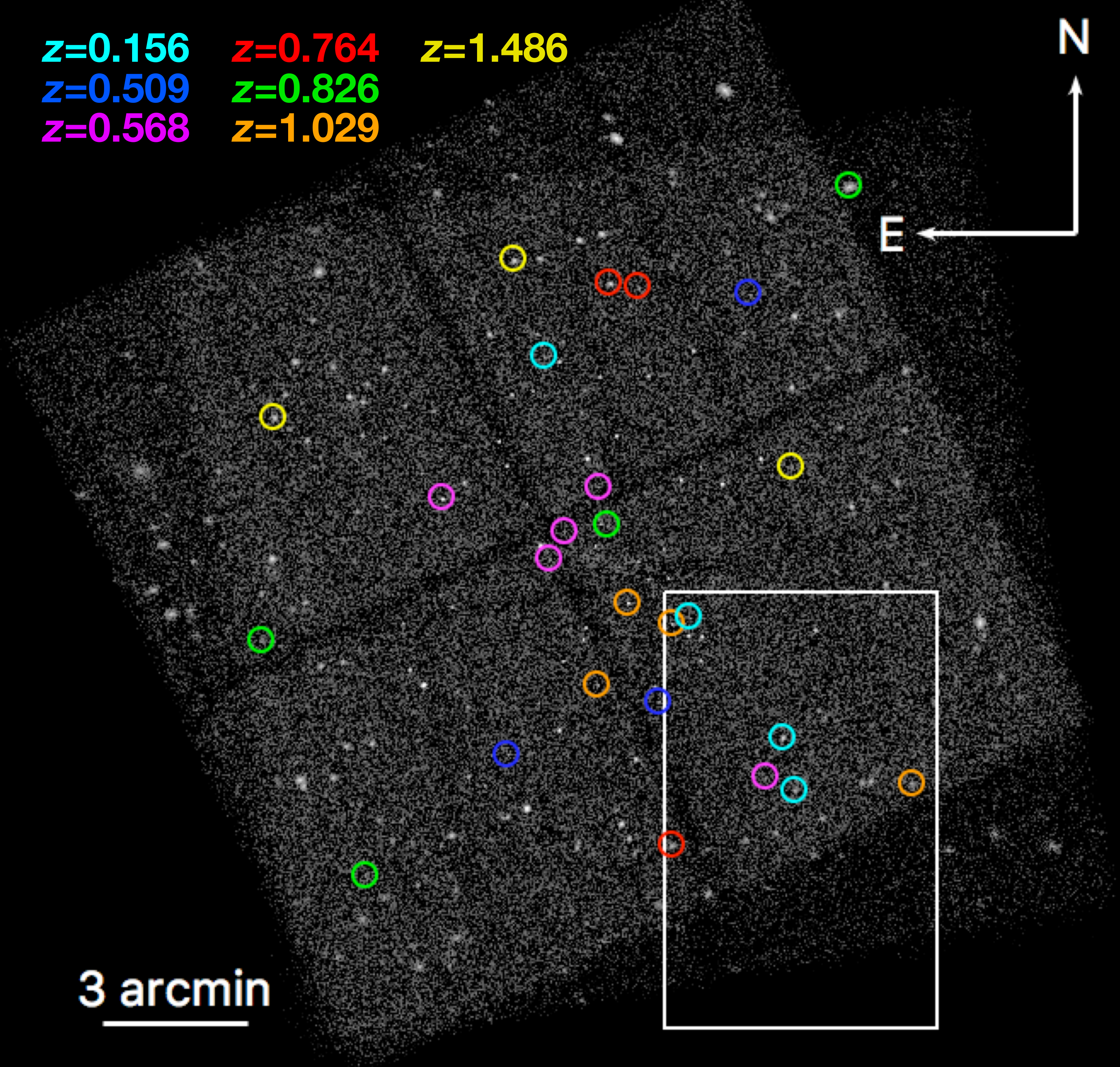} 
 \end{minipage} 
\begin{minipage}[b]{.37\textwidth}
 \centering 
 \includegraphics[width=1.00\textwidth]{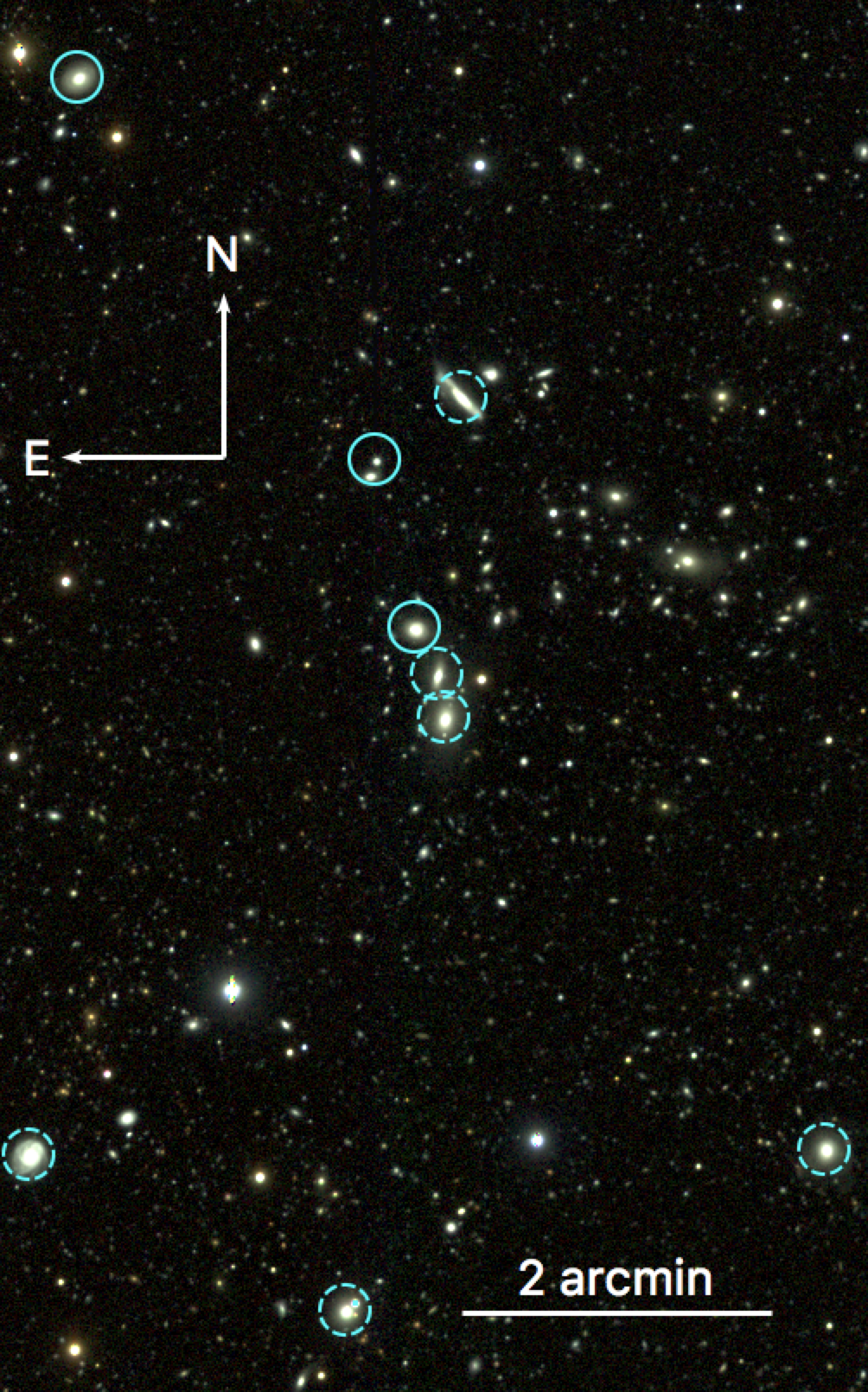} 
 \end{minipage} 
\caption{\normalsize Left: Smoothed 0.5--7\,keV \textit{Chandra} ACIS--I image of the SDSS J1030+0524 field. The X-ray--detected sources which are part of the statistically significant large-scale structures reported in Table \ref{tab:spike_significance} are plotted as 15$^{\prime\prime}$--radius circles: each color corresponds to a different redshift. Right: Tricolor (red: $z$; green: $i$; blue: $r$; all LBT/LBC magnitudes) zoom-in of the $\sim$340$^{\prime\prime}\times$540$^{\prime\prime}$ region of the J1030 field highlighted with a box in the left panel and centered on a $z$=0.156 excess. The X-ray detected (undetected) sources are plotted with a solid (dashed) cyan circle.
}\label{fig:structures_map}
\end{figure*}

Similarly to what was observed in the CDF-S, in J1030 the AGN in the redshift peaks are distributed over the entire \cha\ field of view, on physical scales ranging from 2-3\,Mpc at $z$=0.15 to 5-8\,Mpc at $z$=1.5 (see Figure \ref{fig:structures_map}, left panel). In the CDF-S most of the redshift structures traced with only a few AGN were later confirmed with way larger statistical significance through spectroscopic surveys targeting non-active galaxies \citep[e.g.,][]{popesso09,balestra10}. The same appears to be true for J1030: as shown in Figure \ref{fig:z_vs_cdf}, many of the structures detected in the X-rays can be also recognised in the redshift distribution that includes both the X-ray selected AGNs and normal galaxies in the field for which we derived a spectroscopic redshift by compiling a list of literature redshifts and adding all the filler targets from our spectroscopic programs. As it can be seen in Table~\ref{tab:spike_significance}, in five out of seven cases the significance of the structure increases when adding to the sample the non X-ray detected sources We caution that this second sample is built in a non-uniform and non systematic way, which prevents us from exploring in greater detail the ``X-ray and non X-ray'' structures.

While many sources are distributed over the whole \cha\ field of view, several galaxy members of the identified structures are well localized on the sky, and form groups and clusters on smaller scales. An example is shown in Figure \ref{fig:structures_map}, right panel, where we plot a tri-color ($riz$) LBT/LBC image of a $\sim$350$^{\prime\prime}\times$550$^{\prime\prime}$ region of the J1030 field. Several objects at $z\sim$0.156, both detected and undetected in the X-rays, are found at angular distances of $\lesssim$8$^\prime$, i.e., at a physical distance $d_{\rm L}\lesssim$1.3\,Mpc. 

In summary, similarly to what was already observed in the CDF-S \citep[e.g.,][]{gilli03,silverman10}, we confirm that AGN are excellent tracers of large-scale structures on scales of several Mpc. In particular, the large scale sheets and filaments of matter that intersect at galaxy clusters and groups can be discovered effectively even through observations of a small number of X-ray selected AGN.

Besides the structures discussed above, we also searched for any redshift excess at $z$=1.7, which is the redshift of the Compton-thick Fanaroff-Riley type II (FRII) radio galaxy around which a galaxy overdensity (d$\sim$800\,kpc) was recently discovered by our group combining LBT/LUCI, VLT/MUSE and ALMA observations \citep{gilli19,damato20}. We found another spectroscopically confirmed X-ray source at $z$=1.7 (the BL-AGN XID 16), but the statistical significance of a two-source peak is indeed limited (P=$7\times 10^{-2}$). Furthermore, there are 18 (12) sources which lack a spectroscopic redshift and have photometric redshift in the range $z_{\rm phot}$=[1.5--1.9] ([1.6--1.8]). These are all promising candidates for future spectroscopic campaigns aimed at tracking the large-scale structure over a scale of $\sim$8\,$\times$8\,Mpc$^2$ (i.e., the physical scale of the \cha\ J1030 field at $z=$1.7).

\begingroup
\renewcommand*{\arraystretch}{1.5}
\begin{table*}
\centering
\scalebox{1.}{
\vspace{.1cm}
 \begin{tabular}{c|cc|cc}
 \hline
 \hline
  Redshift  &  $S_{\rm X}$   &  P$_{\rm Poiss(N\geq N_{\rm obj,X})}$ & $S_{\rm All}$   &  P$_{\rm Poiss(N\geq N_{\rm obj,All})}$\\
 \hline
      0.156    & 4  & 2.9$\times$10$^{-4}$ & 11 & 6.2$\times$10$^{-9}$ \\ 
      0.509    & 3  & 3.2$\times$10$^{-2}$ &  9 & 6.4$\times$10$^{-4}$ \\ 
      0.568    & 5  & 9.9$\times$10$^{-4}$ & 11 & 2.8$\times$10$^{-5}$ \\ 
      0.764    & 3  & 2.2$\times$10$^{-2}$ &  3 & 1.3$\times$10$^{-1}$ \\ 
      0.826    & 4  & 6.9$\times$10$^{-3}$ & 10 & 3.3$\times$10$^{-6}$ \\ 
      1.029    & 4  & 2.5$\times$10$^{-2}$ &  7 & 5.5$\times$10$^{-3}$ \\ 
      1.486    & 3  & 6.8$\times$10$^{-3}$ &  4 & 1.5$\times$10$^{-2}$ \\ 
 \hline
 \hline
\end{tabular}}
	\caption{\normalsize Peaks detected in the X-ray detected sources spectroscopic redshift distribution, sorted by increasing redshift. The signal and background distributions are smoothed with $\sigma_{\rm S}$=300\,km\,s$^{-1}$ and $\sigma_{\rm B}$=1.5$\times$10$^4$\,km\,s$^{-1}$, respectively.	``Redshift'' is the central redshift of each peak, $S$ is the number of sources in a peak, P$_{\rm Poiss(N\geq N_{\rm obj})}$ is the Poissonian probability of observing at least N$_{\rm obj}$ sources given the background value. Values marked with $_{\rm X}$ are derived using the X-ray detected sources only, while values marked with $_{\rm All}$ are derived using both the X-ray sources and non-X-ray targets with a spectroscopic redshift.}
\label{tab:spike_significance}
\end{table*}
\endgroup

\section{The \cha\ J1030 high-redshift sample}\label{sec:high-z}
Out of 243 \cha\ J1030 extragalactic sources, 25 have a redshift $z_{\rm best}>$3. Since the spectroscopic campaign initially prioritized optically bright sources, only three of them have a $z_{\rm spec}$, one of which is the $z$=6.3 QSO at the center of the J1030 field. For this reason, in the rest of this section we will not use the photometric redshift nominal values, but rather their probability distribution functions, following an approach adopted in other works on high-$z$ AGN in X-ray surveys \citep[e.g.,][]{marchesi16b,vito18}.  This approach implies that we include in our analysis not only those sources with PDZ peaking at {\it z}$\,>\,$3, but also objects with a nominal photometric redshift $z_{\rm phot}<$3 but with a fraction of PDZ $>$0 at {\it z}$\,>\,$3. 

Following the approach discussed in, for example, \citet{marchesi16b}, we decided to use a conservative approach when analyzing the \cha\ J1030 high-redshift sample.  Consequently, we defined a ``lower reliability'' class of sources, which includes $i$) 7 sources which only have a detection in the nIR IRAC bands; $ii$) 8 sources which only have a detection in both the IRAC bands and in the $K$-band; $iii$) 5 sources that have a detection only in the IRAC bands and in the LBT $r$-Deep catalog; $iv$) One object that has a detection only in the IRAC bands and in the HST H-band. Overall, this ``lower reliability sample'' contains 21 sources, 13 of which have a nominal photometric redshift value $z_{\rm phot}>$3. We note that all these objects have a flat, very poorly constrained PDZ. If these sources are not included in the computation, the overall {\it z}$\,>\,$3 ({\it z}$\,>\,$4) \cha\ J1030 sample contains 22.6 (10.6) sources.

\begin{figure*} 
\begin{minipage}[b]{.49\textwidth} 
 \centering 
 \includegraphics[width=1.\textwidth]{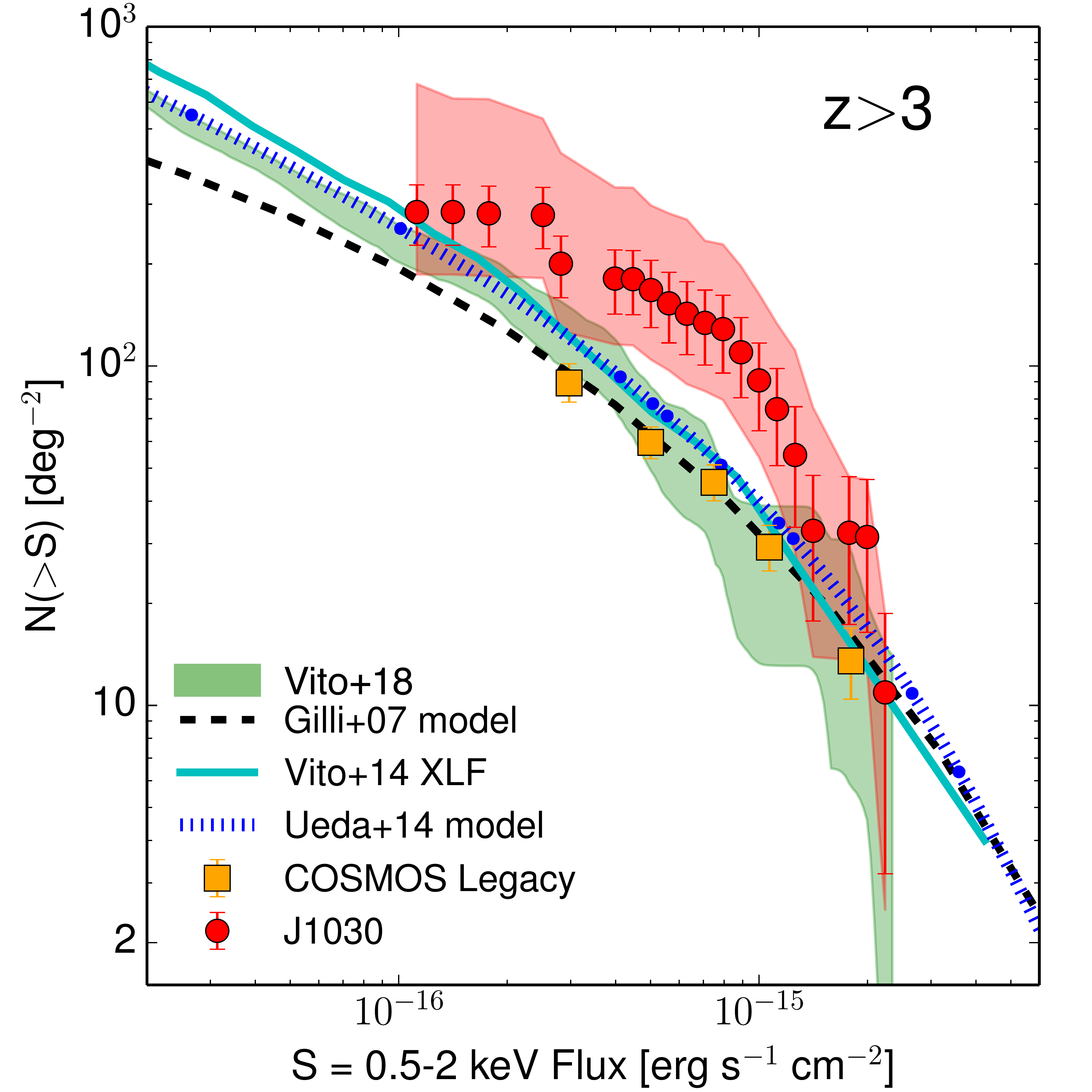} 
 \end{minipage} 
\begin{minipage}[b]{.49\textwidth}
 \centering 
 \includegraphics[width=1.00\textwidth]{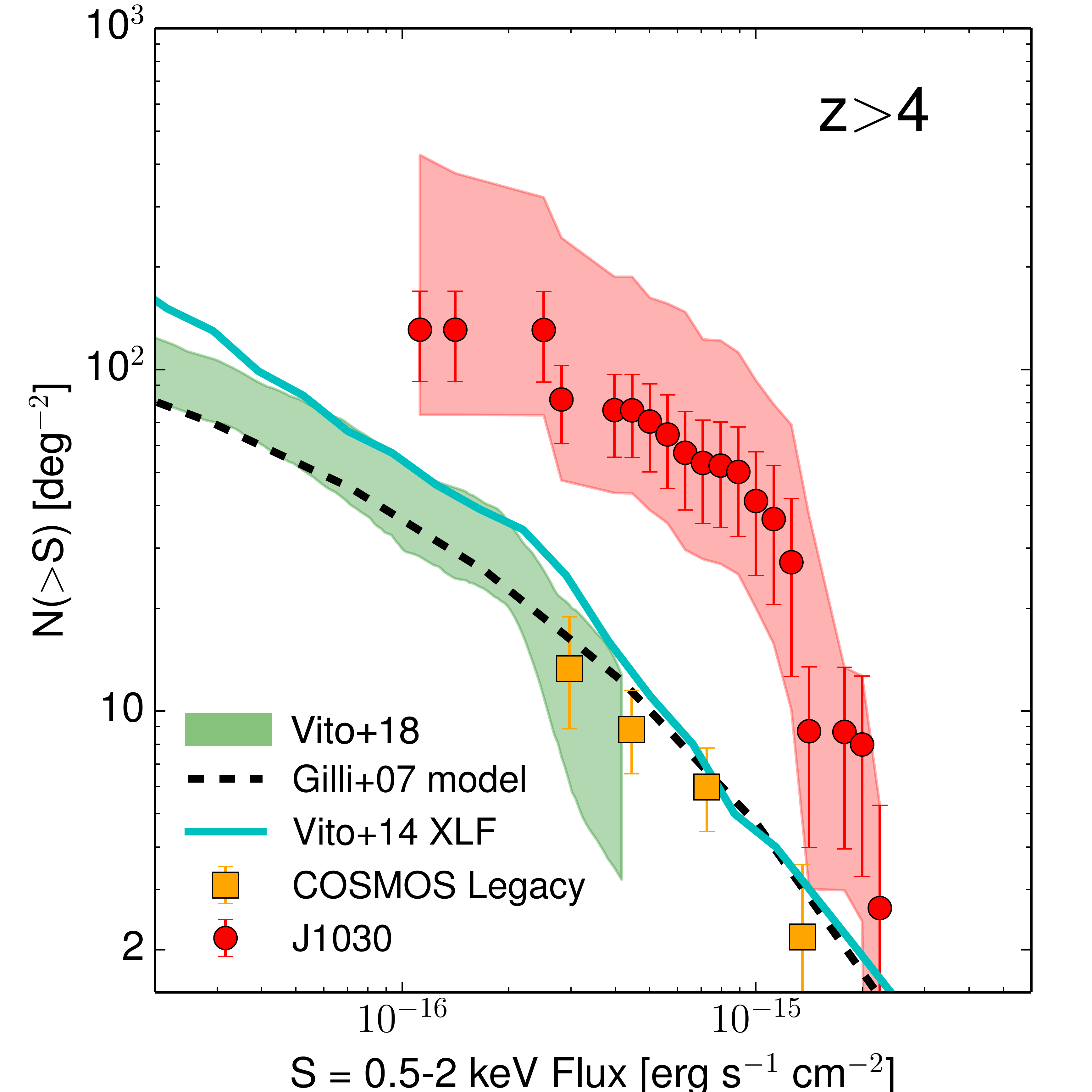} 
 \end{minipage} 
\caption{\normalsize {\it z}$\,>\,$3 (left) and {\it z}$\,>\,$4 (right) number counts for \cha\ J1030 (red circles): the error bars are the Poissonian uncertainties. The upper boundary of the shaded area is obtained including in the computation 21 ``low-reliability'' sources, which are those objects with only a $K$--band and/or an IRAC-band detection, as well as two objects with no counterpart. The lower boundary is computed assuming a conservative PDZ shape (see the text for more details). The number counts obtained  in the same redshift range in the \cha\ COSMOS-Legacy survey \citep[orange squares,][]{marchesi16b} and in the CDF-S and CDF-N surveys \citep[green area][]{vito18} are also shown for comparison, together with the predictions from the \citet[][cyan solid line]{vito14} and \citet[][blue dotted line]{ueda14} X-ray luminosity functions, and those of the \citet{gilli07} AGN population synthesis model.
}\label{fig:logn-logs_zgt3-4}
\end{figure*}

In Figure \ref{fig:logn-logs_zgt3-4} we report the {\it z}$\,>\,$3 (left panel) and {\it z}$\,>\,$4 cumulative number counts in the observed 0.5--2\,keV band for the \cha\ J1030 sources, computed using the standard equation

\begin{equation}
N(>S)=\sum_{i=1}^{N_S}\frac{w_i}{\Omega_i} [deg^{-2}],
\end{equation}

where $N(>S)$ is the number of sources having flux larger than a given flux $S$, $\Omega_i$ is the sky coverage associated with the flux of the $i$th source, $N_S$ is the number of sources above flux $S$, and $w_{i}$ is the weight linked to the PDZ contribution in the range $z_{\rm low}$--$z_{\rm up}$, $w_{i}$=$\frac{\sum_{\rm z_{low}}^{\rm z_{up}}{PDZ(z)}}{\sum_0^7{PDZ(z)}}$, where $z_{\rm low}$=3,4 and $z_{\rm up}$=7. Sources with a spectroscopic redshift have $w_{i}$=1. 

The \cha\ J1030 ``clean'' number counts reported in Figure~\ref{fig:logn-logs_zgt3-4} as red circles are obtained removing from the sample the 21 ``low reliability'' sources mentioned above. These low-reliability sources are instead accounted for computing the upper boundary of the red shaded area: in the computation of this upper boundary we also include the two (out of six) sources lacking a counterpart that are detected in the 0.5--2\,keV band, namely, XID 96 and XID 187. For these two sources, we assume a PDZ($z$) equal to 0 at {\it z}$\,<\,$3, and equal to 5$\times$10$^{-3}$ at $z\geq$3 (so that the sum of all PDZ($z$) values is equal to 1), working under the strong assumption that the two sources are both at {\it z}$\,>\,$3. We note that the contribution of these objects to the boundary is nonetheless fairly marginal, and our results would not significantly change if XID 96 and XID 187 were not included in the computation. The upper boundary of the shaded area is therefore obtained adding to the ``clean'' number counts those of the ``low reliability'' and ``no-counterpart'' subsamples, and further adding the 1\,$\sigma$ Poissonian uncertainty.

Finally, to compute the shaded area lower boundary we assume that the PDZ of our sources can vary over the whole redshift range, without applying any absolute magnitude cut (see Section~\ref{sec:SED_fitting}). This approach affects almost exclusively those objects best fitted with an AGN template, and produces as a result number counts that are $\sim$30\,\% lower than those obtained with the magnitude correction. In the figures, the lower boundary is computed subtracting to these new number counts their 1\,$\sigma$ Poissonian uncertainty.

As a comparison, we also report the high-$z$ AGN counts obtained in the deepest X-ray surveys currently available \citep[the CDF-S 7\,Ms and the CDF-N 2\,Ms, see][]{vito18} and in the \cha\ COSMOS Legacy Survey \citep{marchesi16b}. The number counts derived using the \citet{gilli07} AGN population synthesis model, as well as those from the AGN X-ray luminosity functions by \citet{ueda14} and \citet{vito14} are also shown for comparison. As it can be seen, the \cha\ J1030 number counts lie slightly above the others at {\it z}$\,>\,$3, and the excess with respect to both the predictions of the models and previous observational results becomes even larger (a factor $\sim$4--5) at {\it z}$\,>\,$4.

\begin{figure*} 
\begin{minipage}[b]{.32\textwidth} 
 \centering 
 \includegraphics[width=0.99\textwidth]{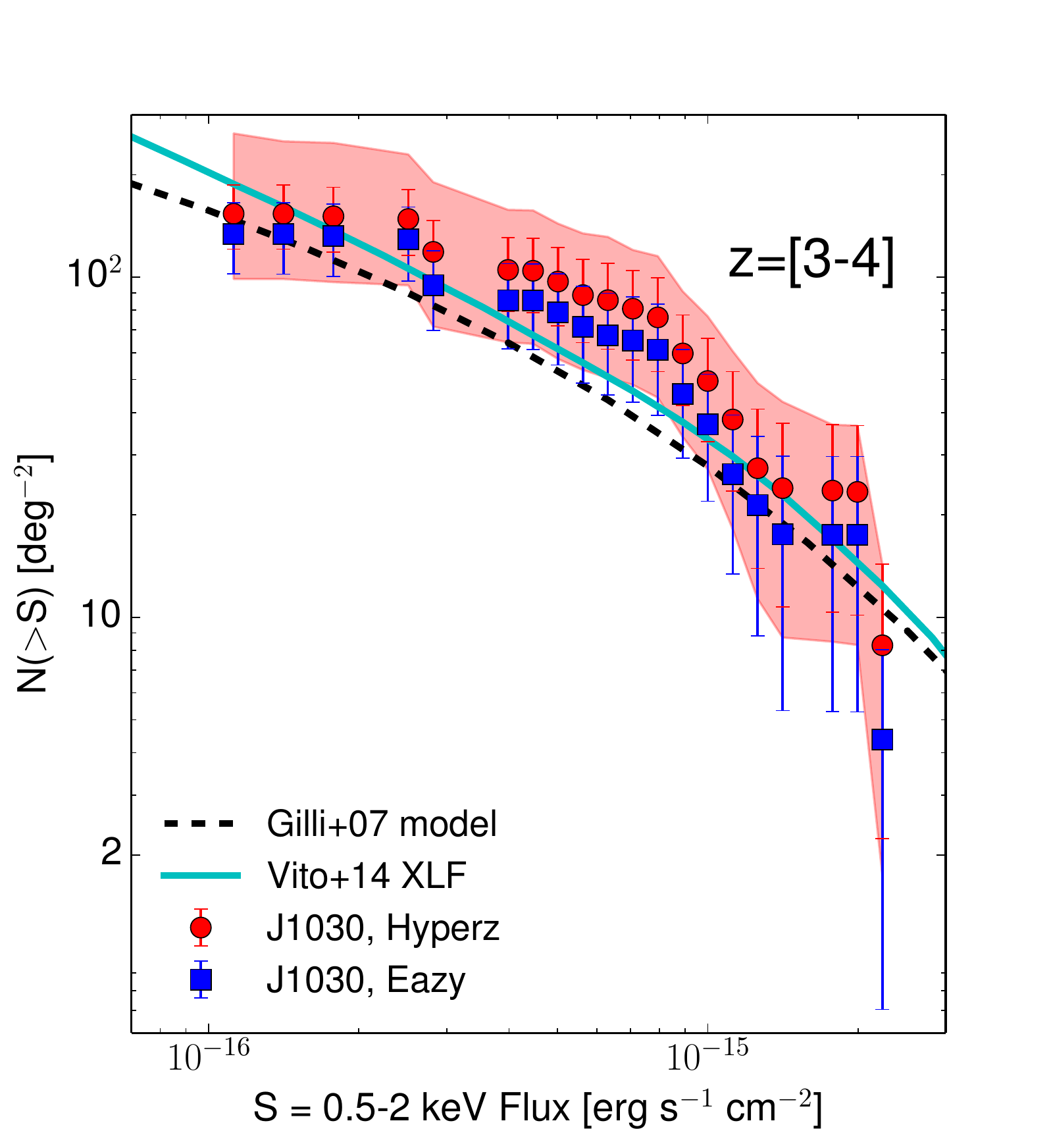} 
 \end{minipage} 
\begin{minipage}[b]{.33\textwidth}
 \centering 
 \includegraphics[width=1.00\textwidth]{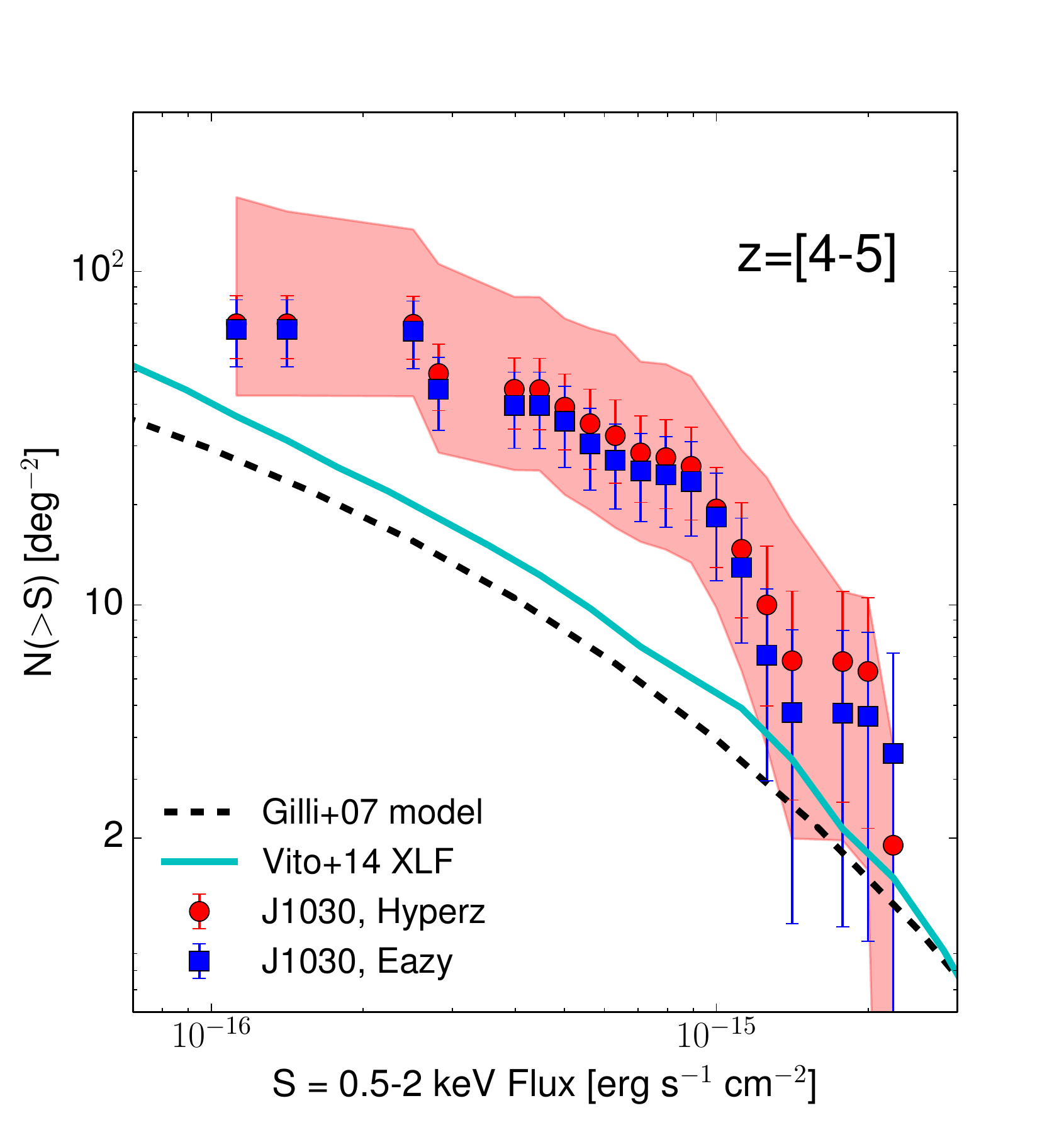} 
 \end{minipage} 
\begin{minipage}[b]{.32\textwidth}
 \centering 
 \includegraphics[width=1.00\textwidth]{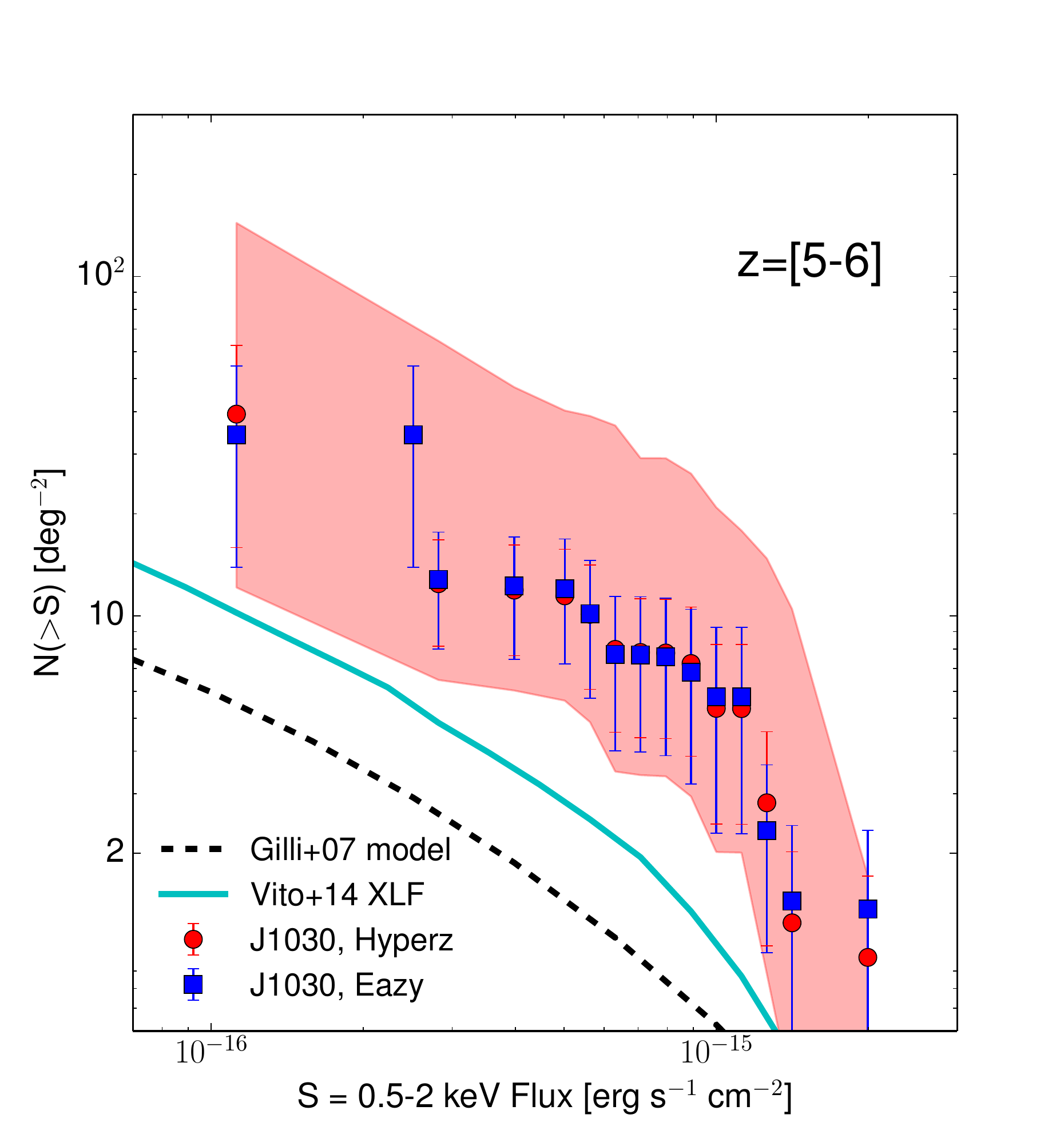} 
\end{minipage}
\caption{\normalsize \cha\ J1030 0.5--2\,keV number counts computed using the \texttt{Hyperz} photometric redshifts (red circles in the redshift bins $z$=[3--4] (left), $z$=[4--5] (center), and $z$=[5--6] (right). The upper boundary of the shaded area is computed including the ``low-reliability'' sources defined in the text and two objects lacking a counterpart. The lower boundary is computed assuming a conservative PDZ shape (see the text for more details). The number counts derived using the \texttt{EAzY} photometric redshifts (blue squares) are also plotted for comparison. 
The number counts in the same redshift ranges derived using the \citet{gilli07} AGN population synthesis model (black dashed line) and the \citet{vito14} XLF (cyan solid line) are also shown for comparison.
}\label{fig:logn-logs_z_bin}
\end{figure*}

To further investigate the origin of this excess, we divide the \cha\ J1030 number counts in three different redshift bins, $z$=[3--4], $z$=[4--5] and $z$=[5--6], which we report in Figure \ref{fig:logn-logs_z_bin}. We note that in doing so we remove the contribution of the $z$=6.3 QSO, whose presence creates, by selection, a strong overdensity at {\it z}$\,>\,$5: indeed, at the 0.5--2\,keV flux of SDSS J1030+0525 one would expect to have $\sim$1 such source per square degree \citep[see, e.g., Figure 8 in][]{marchesi16b} at $z\geq$5, while the J1030 contribution is $N$=12.9 sources/deg$^2$.

In the redshift range $z$=[3--4], we observe a mild excess, which is nonetheless consistent within the uncertainties, with the \citet{vito14} model predictions, and a $\sim$50\,\% excess with respect to the predictions of the \citet{gilli07} model. We note that this $z$=[3--4] excess contributes to most of the excess observed at $z>$3.  
In the redshift range $z$=[4--5] we instead measure a factor $\sim$3$^{+3}_{-1.5}$ excess with respect to the predictions of the \citet{vito14} model\footnote{The upper and lower uncertainties on the ratio are computed using the upper and lower boundaries of the shaded area shown in Figure~\ref{fig:logn-logs_z_bin}.}, and a factor $\sim$4$^{+4}_{-2}$ excess with respect to those of the \citet{gilli07} one. Such an excess is the main cause of the one observed in the {\it z}$\,>\,$4 number counts.  Finally, in the redshift range $z$=[5--6] our data shows an excess by a factor $\sim$3.5$^{+14}_{-2}$ with respect to the \citet{vito14} model predictions, and by a factor $\sim$6$^{+23}_{-4}$ with respect to the \citet{gilli07} model ones. 
In Figure~\ref{fig:logn-logs_z_bin} we also plot the \cha\ J1030 number counts computed using the \texttt{EAzY} photometric redshifts and PDZs (corrected by $a$=3.7, like the \texttt{Hyperz} ones, as discussed in Section~\ref{sec:SED_fitting}). As it can be seen, we find an excellent agreement between the two number counts estimates in the redshift ranges $z$=[4--5] and $z$=[5--6]. This shows that the detection of the $z>$4 excess is code--independent. In the redshift range $z$=[3--4], the \texttt{EAzY} number counts are instead slightly (~20\,\%) smaller than the \texttt{Hyperz} ones, an evidence that partially reduces the significance of the excess.

To better quantify the significance of the observed high-$z$ excess, we convolve the \citealt{gilli07} (\citealt{vito14}) number counts predictions by the \cha\ J1030 sky coverage and we find an expected number of 7, 1.2 e 0.2 (8, 1.3, 0.2) AGN detections in the redshift bins $z$=[3--4], $z$=[4--5] and $z$=[5--6], respectively. In the same redshift bins, the integrated number of \cha\ J1030 sources (without taking into account the ``lower reliability'' ones) is 12.0, 6.2 and 2.6. These numbers show that the excess we measured, although numerically large, might in principle be caused by a small number of objects.

The intriguing excess we measure in our number counts at {\it z}$\,>\,$4 might hint at the existence of an AGN overdensity in the same redshift range. However, we point out that at the moment the \cha\ J1030 sample in the $z$=[4--6] is made only of sources with a photometric redshift. For this reason, our results need to be validated with an extensive spectroscopic campaign in the optical and near-infrared aimed at significantly increasing the {\it z}$\,>\,$3 spectroscopic completeness of our sample (which currently is $f_{\rm spec}$=3/25=12\,\%).

\section{Summary and conclusions}\label{sec:summary}
In this work, we presented the spectroscopic and photometric redshift catalog for the \cha\ J1030 survey, the fifth deepest X-ray survey currently available. We report here the main results of our analysis.

\begin{enumerate}
    \item The \cha\ J1030 sample contains 243 extragalactic sources for which we measured a redshift. Seven sources are X-ray emitting  stars, while other six sources lack a counterpart in any band and therefore a redshift: these latter objects are either spurious X-ray detections, or candidate high-redshift sources.
    \item We measured a spectroscopic redshift for 123 out of 256 \cha\ J1030 objects. The vast majority ($\sim 95\%$) of these spectroscopic redshifts were obtained through a 52 hours INAF--LBT Strategic Program that was granted to our group in 2017.
    \item Out of 123 sources with a spectroscopic redshift, 63 are classified as AGN (43 broad-line AGN; 20 narrow-line AGN), while 60 have spectra dominated by the host galaxy emission. More in detail, 28 sources are emission line galaxies, while 32 are early-type galaxies, a significant fraction of which show faint emission lines, likely linked to obscured and/or weak SMBH accretion.
    \item Taking advantage of the excellent optical-to-nIR coverage of the J1030 field, we computed a photometric redshift for 243 extragalactic sources. We used the spectroscopic subsample to estimate how reliable are the photometric redshifts: we achieved a median accuracy $\sigma_{\rm NMAD}$=1.48$\times$median(||$z_{\rm phot}$-$z_{\rm spec}$||/(1+$z_{\rm spec}$))=0.065, with a fraction of outliers $\eta$=20.5\,\%. Both values are in good agreement with those obtained in other X-ray surveys which computed photometric redshifts using only broad-band photometry data points \citep[e.g., Stripe 82X,][
    Lockman Hole, \citealt{fotopoulou12} ]{ananna17}. Notably, when excluding from the sample the BL-AGN population, whose SED is on average more complex to characterize, the outlier fraction drops to 12.7\,\%. The quality of the photometric redshifts was further validated by computing them with two independent codes, \texttt{Hyperz} and \texttt{EAzY}, and obtaining a good agreement (outlier fraction $<$20\,\%).
    \item Following the approach presented in \citet{gilli03}, we searched for potential large-scale structures in the \cha\ J1030 sample. We found seven different structures in the redshift range $z_{\rm spec}$=[0.15--1.5], with the three most prominent ones being at $z$=0.156, $z$=0.568 and $z$=0.826. 
    \item All the seven structures detected in the \cha\ J1030 sample were confirmed when adding to the spectroscopic sample non-X-ray galaxies from the literature or filler targets of our spectroscopic programs. In five out of seven cases, the structure significance increased when including to the sample the non X-ray sources. This result highlights how  X-ray selected AGN are highly efficient tracers of large-scale structures.
    \item Using our photometric redshifts in a probabilistic way (i.e., assigning to each source a weight based on the fraction of its PDZ in a certain redshift range), we studied the \cha\ J1030 high-redshift sample. We found potential evidence of an overdensity at {\it z}$\,>\,$4: in the redshift range $z$=[4--5] ($z$=[5--6]) we measure a factor $\sim$3 ($\sim$3.5) excess with respect to the predictions of the \citet{vito14} model.
\end{enumerate}

 While the count excess we found at {\it z}$\,>\,$4 would be an intriguing discovery, we point out that our results need to be validated by increasing the \cha\ J1030 spectroscopic completeness at {\it z}$\,>\,$3: currently only 3 out of 25 {\it z}$\,>\,$3 sources have $z_{\rm spec}$, one being the $z$=6.3 QSO itself. For this reason, we aim to target a significant fraction of our {\it z}$\,>\,$3 population with optical/nIR spectrographs at 8--10\,m class telescopes in the near future.

\section*{Acknowledgements}
We thank the anonymous referee for their helpful report, which helped us in significantly improving the paper. We thank Francesca Civano and Eros Vanzella for the useful suggestions. We acknowledge the support from the LBT-Italian Coordination Facility for the execution of observations, data distribution and reduction. The LBT is an international collaboration among institutions in the United States, Italy and Germany. LBT Corporation partners are the University of Arizona on behalf of the Arizona university system; Istituto Nazionale di Astrofisica, Italy; LBT Beteiligungsgesellschaft, Germany, representing the Max-Planck Society, the Astrophysical Institute Potsdam, and Heidelberg University; The Ohio State University, and The Research Corporation, on behalf of The University of Notre Dame, University of Minnesota and University of Virginia. The scientific results reported in this article are partly based on observations made by the \cha\ X-ray Observatory. This work made use of data taken under the ESO program ID 095.A0714(A). We acknowledge financial contribution from the agreement ASI-INAF n. 2017-14-H.O. KI acknowledges support by the Spanish MICINN under grant Proyecto/AEI/10.13039/501100011033 and ``Unit of excellence Mar\'ia de Maeztu 2020-2023'' awarded to ICCUB (CEX2019-000918-M). TM acknowledges the support provided by NASA through grant Nos. HST-GO-15702.002 and HST-AR-15804.002-A from the Space Telescope Science Institute, which is operated by AURA, Inc., under NASA contract NAS 5-26555. We acknowledge support from the National Aeronautics and Space Administration (NASA) under award number NNX16AN49G, issued through the NNH15ZDA001N Astrophysics Data Analysis Program (ADAP).

\bibliographystyle{aa}
\bibliography{J1030_spec_photo-z}

\end{document}